\documentclass[aps,onecolumn,notitlepage,superscriptaddress,floatfix,showkeys]{revtex4-1}
\usepackage{amsmath,amssymb,amsfonts}   
\usepackage{lineno,hyperref}
\usepackage{graphicx}
\usepackage{epstopdf}
\usepackage[]{appendix}
\modulolinenumbers[1]

\renewcommand{\thesection}{\arabic{section}}
\renewcommand{\thesubsection}{\thesection.\arabic{subsection}}
\renewcommand{\thesubsubsection}{\thesubsection.\arabic{subsubsection}}

\makeatletter
\renewcommand{\p@subsection}{}
\renewcommand{\p@subsubsection}{}
\makeatother

\begin{document}


\title{Fast magma ascent, revised estimates from the deglaciation of Iceland}

\author{David W. \surname{Rees Jones}}
\affiliation{University of Cambridge, Bullard Laboratories, Department of Earth Sciences, Madingley Road, Cambridge, CB3 0EZ, UK.}
\affiliation{University of St Andrews, School of Mathematics and Statistics, Mathematical Institute, North Haugh, St Andrews, KY16 9SS, United Kingdom}
\author{John F. \surname{Rudge}}
\affiliation{University of Cambridge, Bullard Laboratories, Department of Earth Sciences, Madingley Road, Cambridge, CB3 0EZ, UK.}

\begin{abstract}
Partial melting of asthenospheric mantle generates magma that supplies volcanic systems. 
The timescale of melt extraction from the mantle has been hotly debated. 
Microstructural measurements of permeability typically suggest relatively slow melt extraction (1~m/yr) whereas geochemical  (Uranium-decay series) and geophysical observations suggest much faster melt extraction (100~m/yr). 
The deglaciation of Iceland triggered additional mantle melting and magma flux at the surface. 
The rapid response has been used to argue for relatively rapid melt extraction. 
However, this episode must, at least to some extent, be unrepresentative, because the rates of magma eruption at the surface increased about thirty-fold relative to the steady state. 
Our goal is to quantify this unrepresentativeness. 
We develop a one-dimensional, time-dependent and nonlinear (far from steady-state), model forced by the most recent, and best mapped, Icelandic deglaciation. 
We find that 30~m/yr is the best estimate of the steady-state maximum melt velocity. 
This is a factor of about 3 smaller than previously claimed, but still relatively fast. 
We translate these estimates to other mid-ocean ridges accounting for differences in passive and active upwelling and degree of melting.
We find that fast melt extraction greater than about 10~m/yr prevails globally.
\end{abstract}


\maketitle

\section{Introduction}

Magma produced in the upper mantle migrates towards the surface where it erupts forming igneous rocks at mid-ocean ridges, ocean islands and  subduction-zone volcanoes. 
However, estimates of the speed at which magma migrates upwards are highly uncertain ranging between $\lesssim 1$~m/yr to $1000$~m/yr \citep{connolly09}. 
Microstructural models and measurements typically yield relatively slow velocity estimates around 1~m/yr, assuming diffuse porous flow \citep{wark03,zhu03,Rudge18}.
By contrast, evidence from the disequilibrium of $^{230}$Th and $^{226}$Ra points to much more rapid magma ascent velocities around 100~m/yr \citep{stracke06,Elliot2014}.
In the arc context, \citet{Turner2001} argued that the observed $^{226}$Ra-excess in arc lavas comes from slab-derived fluids which suggests that the melt takes less than a few hundred years to traverse the entire melting region, equivalent to a velocity around 1000~m/yr.
\citet{rubin05} suggested melt could traverse the melting region over only a few decades (equivalent to a velocity around 10,000~m/yr) based on the disequilibrium of $^{210}$Pb, but the observed disequilibrium is generally attributed to magma degassing \citep{Turner2012}. 
Geophysical observations (seismic and MT) also provide constraints on magma ascent velocities, since
a relatively small residual porosity points to relatively rapid magma velocities, as discussed by \citet{Armitage2019}. 
However, such constraints are rather indirect, going from seismic wavespeed, to porosity, to magma velocity. 
Indeed, the relationship between wavespeed and porosity is poorly constrained.

The deglaciation of Iceland has the potential to resolve aspects of this controversy. 
When Iceland deglaciated at the end of the last ice age, the removal of the ice load and postglacial rebound triggered a burst of magmatic activity.
The timing and volume of magma production over this period have been extensively mapped \cite[and references therein]{Eksinchol2019}.
This record has been used to infer relatively rapid melt ascent velocities \citep{jull96,maclennan02,Eksinchol2019}.
In combination with measurements of the changing REE composition of erupted basalts, the most recent estimate is of a magma velocity of about 100~m/yr.
The effect of the deglaciation of Iceland on mantle melting 
may be important for CO$_2$ outgassing over the past 120~kyr \citep{Armitage2019}
and is thought to have an ongoing influence \citep{Schmidt2013}.
The continued melting of ice both in Iceland and elsewhere could increase future volcanic hazard \citep{Tuffen2010}.

However, in the almost two-decade period since \citet{maclennan02} argued for a magma velocity exceeding 50~m/yr, the subject has still been debated \citep[e.g.][]{connolly09,lund2011,miller14}.
One reason is that Iceland is, at least to some extent, unrepresentative of magmatic processes more widely. 
First, geographically, Iceland is both a rift-zone at the northern end of the mid-Atlantic ridge and is also influenced by a mantle plume \citep{Shorttle2011}.
It is therefore unlike a simple mid-ocean ridge or a simple ocean island. 
Second, temporally, the Icelandic deglaciation saw the rate of magmatism increase ten- to hundred-fold relative to its typical steady-state value \citep{maclennan02,Eksinchol2019}.
Thus it seems plausible that the porosity was considerably higher and magma velocity was considerably faster than usual at this time. 
The goal of this paper is to account properly for the temporarily increased magma velocity and thereby estimate the true steady-state magma velocity beneath Iceland. 
We then use the information from Iceland to estimate magma velocity more generally across other ridge segments.

Glaciation and magmatism are also coupled because changes in sea level affect mid-ocean ridge magmatism and carbon fluxes \citep{huybers2009,lund2011,crowley2015,burley2015}. 
\citet{huybers2017} raised the intriguing possibility that changes in magmatic carbon fluxes could be an important pacing mechanism for Pleistocene climate cycles, potentially explaining the mid-Pleistocene transition between 40~kyr cycles (in line with orbital obliquity forcing) and 100~kyr cycles. 
However, this proposed mechanism depends on the magnitude and time delay (lag) of the magmatic response.
Therefore, \citet{Cerpa2019} calculated these quantities and showed that they are both controlled by the melt ascent velocity. 
For a maximum melt velocity of 1~m/yr, the lag is up to about 20~kyr, which is within the range that \citet{huybers2017} argued were conducive to causing 100~kyr climate cycles.
However, for a maximum melt velocity of 10~m/yr, the lag is up to about 2~kyr, which is outside the range.
Improved estimates of the maximum melt velocity from the deglaciation of Iceland are therefore crucial to assess the viability of this magmatic explanation of 100~kyr Pleistocene climate cycles. 

Methodologically, we also need to go beyond the study of \citet{Cerpa2019}, because the variation in magmatism caused by the deglaciation of Iceland was vastly greater than that caused by changes in sea level. 
Iceland lost about 2~km of ice in about 1~kyr while global sea level varied by about 100~m over periods of 40--100~kyr \citep{siddall2010}.
Thus, unlike in Iceland, the mid-ocean ridge magma flux only deviates from the steady-state by in the region of 10\% \citep{Cerpa2019}.
This allowed \citet{Cerpa2019} to make a \textit{linear} approximation, in which the system was close to steady state.
In this study, the system is far from steady state, and we develop a methodology capable of handling the full  \textit{nonlinear} time-dependence of the magmatic system. 

\section{Methods}
Beneath Iceland the mantle upwells due to plate spreading and plume buoyancy, causing decompression melting. 
The resulting magma is buoyant relative to the solid and so migrates upwards. 
We use a dynamical model that accounts for mass, momentum and energy conservation based on recent work by \citet{Cerpa2019} and the underlying formulation of \citet{mckenzie84}.
We account for a time-dependent melting rate, generalizing previous steady-state melting column models \citep{Ribe1985,hewitt10}.
The model is similar to the one-dimensional model of \citet{Armitage2019}, except that we do not consider the role of carbon-induced melting or calculate the carbon concentration.
This is reasonable for our purposes, because incompatible volatiles such as carbon have only a modest effect on the melt flux at the surface \citep{Cerpa2019}, although it is important to consider this effect when calculating the carbon flux \citep{Armitage2019}.
Other model differences and limitations are discussed in section~\ref{sec:limitations}.

\subsection{Dynamical model}
We now describe our one-dimensional `melting-column' model. 
The column has height $H$ and the mantle upwelling rate at the base of the column is $W_0$. 
In a one-dimensional model, $W_0$ is also the bulk upwelling rate (of mantle plus magma), which is constant throughout the melting column. 
The increase in the liquid magma flux upwards is offset by the decrease in the solid mantle upwelling upwards \citep{Ribe1985,hewitt10}. 

Mass conservation in the liquid phase (magma) is expressed by
\begin{equation}
\frac{\partial \phi}{\partial t} + \frac{\partial Q}{\partial z} = \Gamma.
\end{equation}
Thus the porosity $\phi$ evolves in time $t$ due to the melting rate $\Gamma$ and the divergence of the liquid flux $Q$, where $z$ is height in the melting column measured from 0 at the bottom to $H$ at the top.
Note that $Q=\phi w$ where $w$ is the magma velocity, the quantity we ultimately wish to infer.
The permeability  $k \phi^n$ of the porous rock increases with porosity with an exponent $n$.
This power-law relationship between porosity and permeability is a semi-empirical law that has been widely used to fit laboratory measurements and microstructural models, as discussed by \citet{miller14}.
Experimental observations point to $n=2.6 \pm 0.2$ \citep{miller14}; however, these experiments are limited to relatively large porosity (greater than 1\%).
\citet{Rudge18} used microstructural models over a wide range of porosities.
These calculations show that $n=2$ in the limit of small porosity, as predicted by \citet{vonbargen86}.
At higher porosity (in the range accessible experimentally), the apparent exponent is higher, although the calculations do not follow an exact power law.
Thus throughout this study we take $n=2$, because in the mantle we expect the porosity to be small, below the experimental range.
The liquid melt rises buoyantly because it is less dense than the solid rock. 
The liquid flux is proportional to this driving buoyancy difference and the permeability.
It is inversely proportional to the liquid viscosity. 
Thus the liquid flux satisfies
\begin{equation}
Q = Q_0 \phi^n,
\end{equation}
where $Q_0 = \Delta \rho g k/\mu$ is a prefactor including the density difference between solid and liquid phase $\Delta \rho$ , gravity $g$, and liquid viscosity $\mu$.  
This is a form of Darcy's law (momentum conservation) appropriate when the porosity and compaction length are small \citep{mckenzie84,Cerpa2019}.
The compaction length is a physical property of a partially molten rock that controls the length scale over which isotropic viscous stresses are communicated \citep{mckenzie84}.
When the compaction length is small, compaction-driven flow can be neglected and the flow is driven by buoyancy alone.

Finally, energy conservation determines the melting rate
\begin{equation}
\Gamma = \Gamma_0 \left[ 1 + \mathcal{A} f(t)\right]
\end{equation}
where $\Gamma_0$ is the steady-state rate of decompression melting and $\mathcal{A}$ is the amplification factor by which it is enhanced during a deglaciation event, which lasts from time $t=0$ to $t=t_d$. 
Thus $f(t) = 1$ within this time window and $f(t) = 0$ outside it. 

\subsection{Steady-state behaviour}
The steady-state solution to this system of equations determines the behaviour of the system prior to the deglaciation event.
Neglecting all the time-dependent terms, we find steady-state solutions (denoted with an overline)
\begin{equation}
\overline{Q} =  \Gamma_0 z, \qquad
\overline{\phi} =  \left(\Gamma_0 z / Q_0 \right)^{1/n}.
\end{equation}
We evaluate these quantities at the top of the melting column where they are maximum:
\begin{equation}
\overline{Q}_\mathrm{max} =  \Gamma_0 H, \qquad
\overline{\phi}_\mathrm{max} =  \left(\Gamma_0 H / Q_0 \right)^{1/n}.
\end{equation}
Note that $\overline{Q}_\mathrm{max}$ can be related to the maximum degree of melting by $F_\mathrm{max} = \overline{Q}_\mathrm{max}/W_0$ \citep{Ribe1985}. 
This allows us to estimate \mbox{$\Gamma_0 = F_\mathrm{max} W_0 / H$}. 
The maximum porosity $\overline{\phi}_\mathrm{max}$ is a balance between melt production and melt extraction, so faster melt extraction (higher $Q_0$) corresponds to a lower maximum porosity.

A key goal of this paper is to estimate the steady-state melt ascent velocity, which is given by
\begin{equation} \label{eq:wbar}
\overline{w} =  \overline{Q}/\overline{\phi} = { \left(\Gamma_0 z \right)^{\frac{n-1}{n}}}{Q_0^{\frac{1}{n}}}.
\end{equation}
Thus the maximum melt velocity (at the top of the column $z=H$) is
\begin{eqnarray}
  \overline{w}_\mathrm{max} & = &   { \left(\Gamma_0 H \right)^{\frac{n-1}{n}}}{Q_0^{\frac{1}{n}}}, \nonumber \\
& = &  { \left(F_\mathrm{max} W_0 \right)^{\frac{n-1}{n}}}{Q_0^{\frac{1}{n}}}. \label{eq:wmax}
\end{eqnarray}
The travel time $T$ for a parcel of melt across the total depth of the melting region is
\begin{equation} \label{eq:travel-time}
T = n H /   \overline{w}_\mathrm{max}  = n H^{\frac{1}{n}}  { \Gamma_0^{\frac{1-n}{n}}}{Q_0^{-\frac{1}{n}}},
\end{equation}
where the factor of $n$ comes from the depth-dependence of the velocity in equation~(\ref{eq:wbar}).

\subsection{Transient behaviour during and after a deglaciation event}
We next calculate the additional magma flux due to deglaciation, which can be compared to field observations.
It is instructive to introduce a scaled set of variables to quantify the increased magma flux and porosity due to a deglaciation event. 
We define
\begin{equation} \label{eq:scaling}
\hat{t}=\frac{t}{\tau}, \quad \hat{z}=\frac{z}{H}, \quad \hat{Q}=\frac{Q-\overline{Q}}{\mathcal{A} \overline{Q}_\mathrm{max}}, \quad \hat{\phi}=\frac{\phi-\overline{\phi}}{\mathcal{A} \overline{\phi}_\mathrm{max}},\end{equation}
where $\tau = H/ \overline{w}_\mathrm{max} \equiv T/n $ is a scaling for time, chosen such that a small porosity perturbation traverses the melting column in scaled time $\hat{t}=1$. 
Thus a parcel of melt, traveling at the steady-state velocity, traverses the melting column in scaled time $\hat{t}=n$, consistent with the result that the velocity of linear porosity waves is $n$ times the melt velocity \citep{scott84,spiegelman93a}.  
In this scaling, $\hat{Q}$ and $\hat{\phi}$ are fractional measures of the extra (i.e. non-steady-state) magma flux and porosity, respectively, per unit of extra melting due to the deglaciation $\mathcal{A}$.

The rescaled set of equations is
\begin{equation} \label{eq:transient-scaled}
\frac{\partial \hat{\phi}}{\partial \hat{t}} + \frac{\partial \hat{Q} }{\partial \hat{z}} = f(\hat{t}) ,
\end{equation}
and
\begin{equation} \label{eq:Q-phi-transient}
\hat{Q} = \frac{Q_0}{\mathcal{A} \overline{Q}_\mathrm{max}} \left [ \left( \overline{\phi} + \mathcal{A} \overline{\phi}_\mathrm{max} \hat{\phi} \right)^n  \, - \, \overline{\phi}^n \right].
\end{equation}
When $n=2$, the latter expression~(\ref{eq:Q-phi-transient}) simplifies to
\begin{equation} \label{eq:Q-phi-transient-n2}
\hat{Q} = 2 \hat{z}^{1/2} \hat{\phi} + \mathcal{A} \hat{\phi}^2.
\end{equation}
The first term on the right-hand-side of equation (\ref{eq:Q-phi-transient-n2}) is the linear part of the enhanced flux caused by elevated melting and porosity from a deglaciation event.
The dependence on $2 \hat{z}^{1/2}$ reflects the depth-dependence of the steady-state melt velocity.
The second term is the nonlinear part of the enhanced flux and is a crucial way this model extends that of \citet{Cerpa2019}.
Throughout this article, we refer to models that include this second term as \textit{nonlinear} and those that neglect it as \textit{linear}.
Linear models are labelled $\mathcal{A} \ll 1$, meaning that the nonlinear amplification factor $\mathcal{A}$ is small.
Physically, a nonlinear model includes the feedback that arises from the elevated melting rate, porosity and melt velocity during deglaciation while a linear model does not include this feedback.
Previous models used to infer melt velocity were linear in this sense \citep{maclennan02,Eksinchol2019}.

We introduce some further definitions; we define 
\begin{equation}
\lambda  = t_d/\tau, 
\end{equation}
which is the dimensionless deglaciation time. 
Thus deglaciation $f(\hat{t})=1$ occurs when \mbox{$0\leq \hat{t} \leq \lambda$}.
The total extra emissions (magmatic output at the top of the column) caused by deglaciation are equal to $\lambda$.
This can be seen by integrating equation~(\ref{eq:transient-scaled}).
Thus we define the relative additional cumulative emissions
\begin{equation}
\hat{F}(\hat{t})  =\lambda^{-1} \int_0^{\hat{t}} \hat{Q}(\hat{z}=1,\hat{t}') \, d\hat{t}'.
\end{equation}

This system of equations can be solved either numerically or analytically using the method of characteristics. 
The latter method is better in that we find a closed-form solution displayed in \ref{app:solution} and derived in the Supplementary Material.
Note that the decomposition into steady-state and time-dependent contributions is not fundamental, both numerical and analytical methods can also be used to solve the full nonlinear equations prior to this decomposition. 

\section{Results} \label{sec:results}
\subsection{History of extra emissions caused by a deglaciation event} \label{sec:timeseries}
\begin{figure}
\noindent\includegraphics[width=0.85\linewidth]{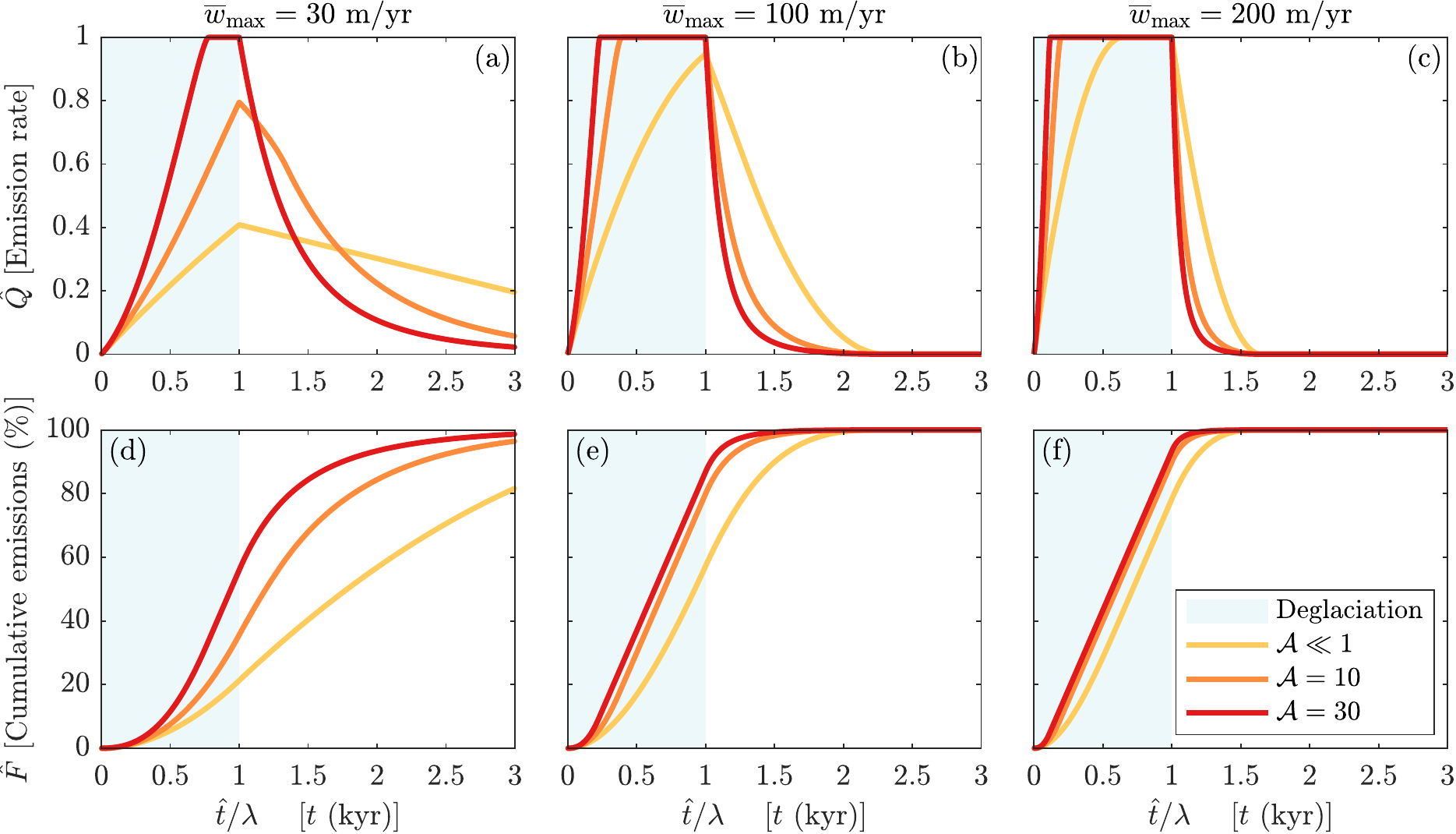}
\caption{Time series of extra volcanic emissions caused by a deglaciation event lasting 1~kyr. 
We show calculations for the linear model ($\mathcal{A}\ll1$) and two nonlinear models ($\mathcal{A} = 10$, $\mathcal{A}= 30$).
First row (a--c) shows the rate of extra volcanic emissions. Second row (d--f) shows the cumulative extra emissions as a percentage of the total. The columns show three possible steady-state maximum melt velocities: $\overline{w}_\mathrm{max}=30$~m/yr (left); $\overline{w}_\mathrm{max}=100$~m/yr (middle); $\overline{w}_\mathrm{max}=200$~m/yr (right).  For this event, the time axis equivalently shows dimensional time in kyr or scaled time $\hat{t}/\lambda$.  }
\label{fig:timeseries}
\end{figure}

Figure \ref{fig:timeseries} shows the evolution of the extra magmatic flux caused by a deglaciation event lasting 1~kyr.
This is roughly the length of the most recent deglaciation after the Younger Dryas.
We compare the linear model ($\mathcal{A}\ll1$) to nonlinear models ($\mathcal{A} = 10$, $\mathcal{A}= 30$).
We show results for a steady-state maximum magma velocity $\overline{w}_\mathrm{max}=30$, $100$ and $200$~m/yr, motivated by recent estimates \citep{Eksinchol2019}.
The velocities shown correspond to steady-state travel times of $4.3$, $1.3$ and $0.65$~kyr, so either slightly longer or slightly shorter than the deglaciation period.
This range therefore covers the main categories of possible results (\ref{app:solution}).

The magmatic emissions rise during the deglaciation period. 
The drop in pressure during this period triggers extra melting throughout the melting column. 
However, these extra melts must migrate to the top of the melting column before they can erupt. 
The rate at which emissions rise therefore reflects the speed of melt transport, with faster melt transport leading to a faster rise in emissions. 
Emissions peak either at the end of the deglaciation period (e.g., for the slowest melt velocities, shown in the yellow curve in panel a) or at an earlier time. 
This earlier time corresponds to the length of time it takes for a parcel of melt to traverse the entire melting region. 
After this time, the extra emissions saturate because the total amount of extra melt (generated over the entire melting region) is the same. 
Beyond the end of the deglaciation period, the extra emissions drop away over a period that is determined by the melt velocity. 

These results are sensitive to the amount of extra melting caused by the deglaciation $\mathcal{A}$ and hence the degree of nonlinearity. 
If the extra melting is negligible ($\mathcal{A} \ll 1$), the time-dependent melt velocity is extremely close to the steady-state melt velocity.  
However, if the extra melting is more considerable ($\mathcal{A} \gg 1$), then the time-dependent melt velocity is considerably faster than at steady state.   
For the most recent deglaciation of Iceland, $\mathcal{A} \approx 30$ is a reasonable estimate \citep{maclennan02,Eksinchol2019}.
This means that emissions peak at some time after the start of deglaciation, rather than continuing to increase slowly over the whole 1~kyr deglaciation period (figure \ref{fig:timeseries}a,b). 
If a faster steady-state maximum melt velocity is used, the differences between linear and nonlinear models are still significant, but slightly smaller because the extra emissions saturate in any case (figure \ref{fig:timeseries}c).

Geological observations of the cumulative emission history are more reliable than estimates of the rate of emissions. 
By integrating the emission history, we can calculate cumulative extra emissions as a percentage of the total (figure \ref{fig:timeseries}d--f). 
The results are again sensitive to the steady-state melt velocity, rising faster with a faster melt velocity. 
Accounting for the extra melting and hence faster effective melt velocity due to deglaciation causes cumulative emissions to rise faster than they would based on the steady-state melt velocity alone.

\subsection{Nonlinear correction to account for effect of extra melting and higher melt velocities caused by deglaciation} \label{sec:NL_correction}
As we saw in the previous section, accounting for elevated melt velocities caused by a deglaciation event leads us to predict a faster rise in emissions than we would infer on the basis of accounting for the steady-state melt velocity alone. 
This suggests that previous studies have overestimated the steady-state melt velocity, because they are based on linear models in which the time-dependent melt velocities is not accounted for.
To correct for this \textit{nonlinear} far from steady-state behaviour (i.e. the full solution of the equations), the true steady-state melt velocity, which we denote $\overline{w}_\mathrm{nonlinear}$, must be slower than the velocity previous studies infer from a \textit{linear} calculation, which we denote $\overline{w}_\mathrm{linear}$.
We define the correction factor as the ratio between the nonlinear and linear velocity estimates.

\begin{figure}
\centering\noindent\includegraphics[width=0.425\linewidth]{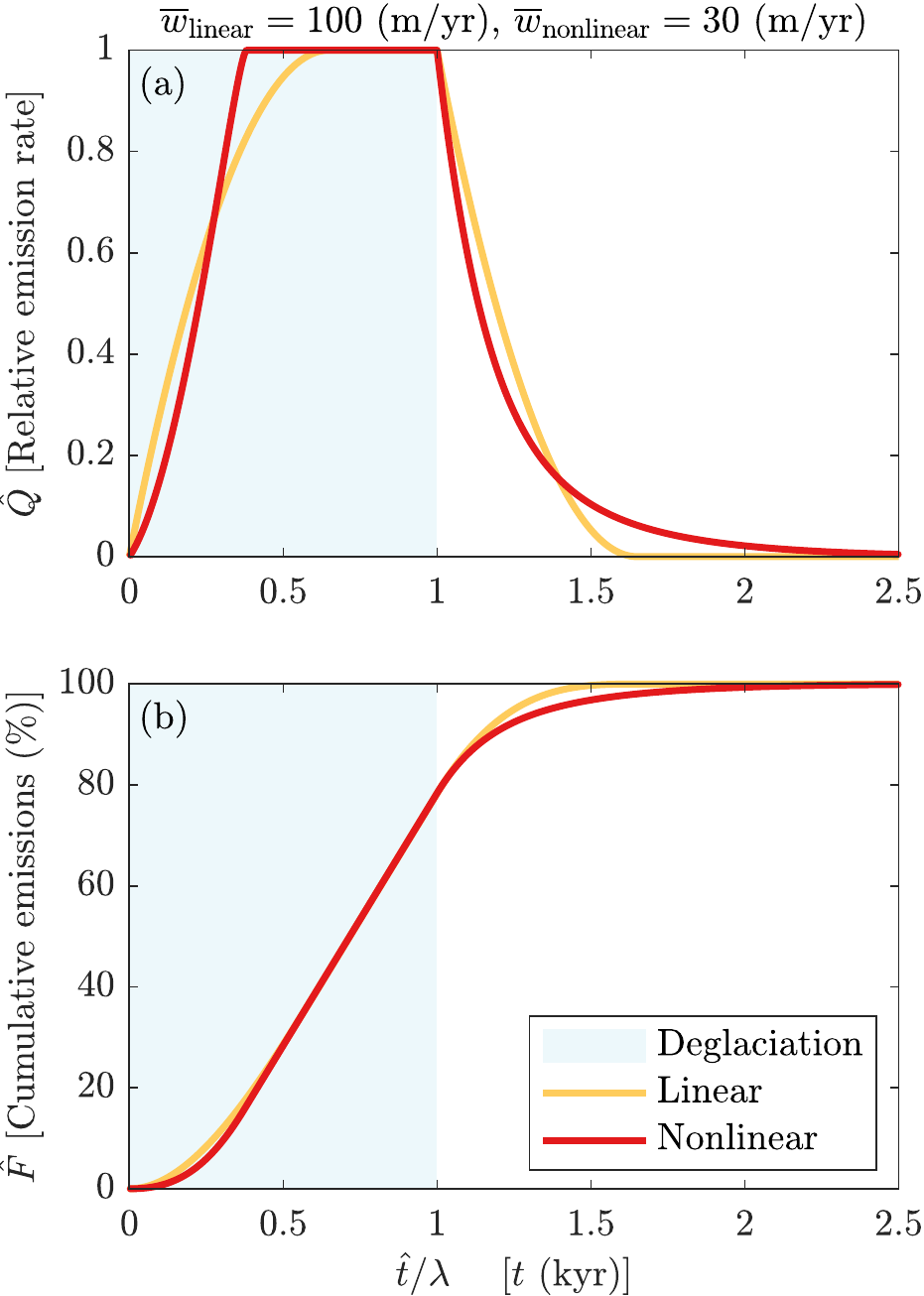}
\caption{Example showing how a fully nonlinear calculation ($\mathcal{A}=30$) has a similar emissions history to a linear calculation ($\mathcal{A} \ll 1$) with a faster steady-state maximum melt velocity (30 and 100~m/yr, respectively). For other details, see caption of figure~\ref{fig:timeseries}.}
\label{fig:lambda_equiv_example}
\end{figure}

Figure~\ref{fig:lambda_equiv_example} illustrates how to correct for this nonlinear far from steady-state behaviour.
We first calculate the emissions history for a nonlinear model with $\overline{w}_\mathrm{nonlinear} = 30$~m/yr and $\mathcal{A} = 30$.
We then calculate the emissions history for a series of linear models with a range of steady-state melt velocities.
We say that an equivalent model is one that takes the same time to reach a certain proportion of total emissions (we use $80$\% in this example).
This definition (based on cumulative emissions) is easiest to compare to observations. 
For this example, $\overline{w}_\mathrm{linear} = 100$~m/yr is an equivalent model.
Thus, if on the basis of a linear model, the estimate of steady-state maximum melt velocity is $\overline{w}_\mathrm{linear} = 100$~m/yr, a better (nonlinear) estimate is $\overline{w}_\mathrm{nonlinear} = 30$~m/yr, and the correction factor is $0.3$. 

\begin{figure}
\centering\noindent\includegraphics[width=0.85\linewidth]{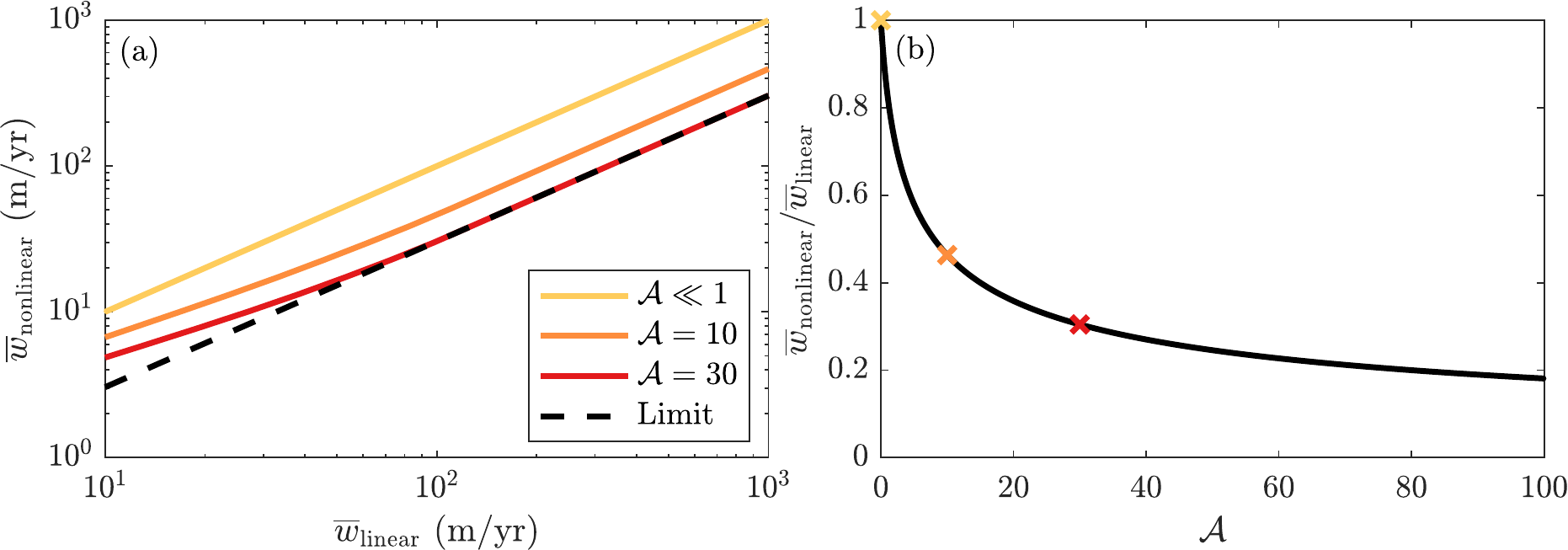}
\caption{(a) Fully nonlinear estimate of maximum melt velocity at steady state as a function of the linear maximum velocity estimate. We show results at various values of $\mathcal{A}$, the additional melting rate factor (nonlinear amplification factor). At large velocities, there is a simple estimate~(\ref{eq:correction_factor}) for the correction factor plotted as the dashed black line. (b) The correction factor given by equation~(\ref{eq:correction_factor}) as function of $\mathcal{A}$. Symbols correspond to curves in panel (a).  }
\label{fig:lambda_equiv}
\end{figure}

Figure~\ref{fig:lambda_equiv} shows the dependence of the correction factor on $\mathcal{A}$, the amount of extra melting caused by deglaciation (nonlinear amplification factor). 
Panel (a) shows an equivalent nonlinear maximum velocity estimate as a function of the linear estimate.
When $\mathcal{A} \ll 1$, the velocities are equal. 
When $\mathcal{A} \gg 1$, the nonlinear maximum velocity estimate can be substantially lower than the linear estimate, so previous studies are likely to overestimate melt velocity. 
If the maximum melt velocity is sufficiently large, we can use analytical expressions for the cumulative emissions (see~\ref{app:cumulative}) to determine a simple expression for the correction factor,
\begin{equation} \label{eq:correction_factor}
\frac{ \overline{w}_\mathrm{nonlinear} }{ \overline{w}_\mathrm{linear} } = \frac{2}{\mathcal{A}} \left( \sqrt{1 + \mathcal{A}} -1 \right) .
\end{equation}
Note that this formula only applies when 80\% of emissions (i.e. the critical fraction we chose) occur during the deglaciation period. 
If not, panel (a) shows graphically the more general behaviour (there are also general equations in the Supplementary Material).
Panel (b) plots equation~(\ref{eq:correction_factor}). 
When $\mathcal{A}$ approaches 0, the correction factor approaches 1, because the nonlinear and linear estimates are the same.
When $\mathcal{A}$ is larger, the correction factor decreases, being in the range of about 0.3---0.5 for $10 < \mathcal{A} < 30$.

While the correction factor is significant, it is noteworthy that it is much smaller than $\mathcal{A}$.
That is, the correction factor is much smaller than the nonlinearity that causes the correction.
The comparatively small correction factor occurs because, for the parameters considered, the extra emissions tend to saturate (figure~\ref{fig:timeseries}).
So accounting for the nonlinearity means that emissions saturate earlier, but this only leads to a modest change in cumulative emissions.

\section{Discussion} \label{sec:discussion}
\subsection{Application to Icelandic melt velocities} \label{sec:discussion_iceland}
\citet{maclennan02} argued that melt velocities beneath Iceland exceeded 50~m/yr based on the timing of the extra emissions that occurred during the most recent deglaciation event.
\citet{Eksinchol2019} argued that a melt velocity of 30~m/yr or less is inconsistent with the most recent emissions history. 
They argued that fast melt velocities (of order 1000~m/yr) are excluded on the basis of observations of incompatible species (Lanthanum) in erupted lavas.
Their preferred estimate of melt velocity is 100~m/yr, consistent with both the  emissions history and Lanthanum measurements.

In this section, we estimate the extent to which these estimates are confounded by the fact that the period of deglaciation was not in steady state.
Given that claims of relatively rapid melt velocity have been most controversial, we give not only `best estimates' but also `lower bounds' on the steady-state melt velocity.

First, we need to translate the previous estimates into our modelling framework. 
The previous studies did not account for the depth-dependence of the melt velocity. 
Thus the estimates they give are arguably more consistent with a travel-time average velocity than the maximum velocity at the top of the column.
The maximum velocity exceeds the average by a factor of $n=2$ (see equation~\ref{eq:travel-time}). 
So the lower bound on the maximum melt velocity from \citet{maclennan02} is about 100~m/yr and a preferred estimate from \citet{Eksinchol2019} is about 200~m/yr.

Second, we need to estimate the additional melting factor $\mathcal{A}$, which controls how far we are from the steady-state and hence the magnitude of the correction factor. 
A lower estimate is $\mathcal{A} = 10$ \citep{Eksinchol2019} and an upper estimate is $\mathcal{A} = 100$ \citep{maclennan02}.
Thus a reasonable intermediate estimate is $\mathcal{A} = 30$.
For this intermediate estimate, the nonlinear correction factor is about 0.3. 
This gives a preferred estimate of the steady-state maximum velocity of 60~m/yr (which is an average velocity of 30~m/yr). 
A lower bound can be found by combining the lowest previous estimate of velocity with the highest estimate for $\mathcal{A}$, giving a maximum steady-state velocity of about 20~m/yr (which is an average velocity of 10~m/yr).

\subsection{Application to other mid-ocean ridge settings} \label{sec:discussion_mor}
Iceland is geologically unusual relative to a normal mid-ocean ridge.
The spreading rate is one of the slowest on the planet and it is influenced by a mantle plume.
In this section, we translate our estimates of maximum melt velocity from Iceland to other mid-ocean ridge settings.

We use equation~(\ref{eq:wmax}) to estimate maximum melt velocity as follows
\begin{equation} \label{eq:estimate1}
\overline{w}_\mathrm{max} = \left( \frac{Q_0} {Q_0^I} \right)^{1/2}  \left( \frac{F_\mathrm{max} W_0} {F_\mathrm{max}^I W_0^I} \right)^{1/2} \overline{w}_\mathrm{max}^I,
\end{equation}
where we use a superscript $^I$ to denote a quantity estimated for Iceland. 
The definition of $Q_0 = \Delta \rho g k/\mu$ makes it likely that $Q_0$ does not vary much geographically (so $Q_0=Q_0^I$) given that $k$ is a property of the mantle rock (the porosity-dependence of permeability is accounted for in the exponent $n=2$). 
It is possible that the melt viscosity $\mu$ (and to a much lesser extent the melt--solid  density difference $\Delta \rho$) might vary as a result of temperature or chemical differences, such as volatile content, particularly in arc settings.
However, variations in volatile content along ridge settings are probably not significant enough to make this a major consideration. 
Instead, we focus on variations in $W_0$ and $F_\mathrm{max}$.

First, the mantle upwelling rate $W_0$ at mid-ocean ridges is generally proportional to the plate spreading rate, assuming plate-driven flow (sometimes called `passive upwelling').
In Iceland, it is possible that the upwelling could be enhanced by additional `active upwelling' associated with the plume buoyancy.

One end-member assumption is that the active upwelling is relatively small \cite[and references therein]{Schmidt2013}.
In this limit, neglecting `active upwelling' only serves to make Iceland appear more different relative to other ridges, since a small amount of active upwelling would partly offset the unusually low spreading rate.  
Thus equation~(\ref{eq:estimate1}) can be simplified to
\begin{equation} \label{eq:estimate2}
\overline{w}_\mathrm{max} = \sqrt{ \frac{F_\mathrm{max} } {F_\mathrm{max}^I}   } \sqrt{  \frac{ U_0} { U_0^I} } \overline{w}_\mathrm{max}^I,
\end{equation}
where $U_0$ is the spreading rate.
The fastest spreading ridges (such as the East-Pacific-Rise) have a spreading rate about 7 times faster than Iceland.
Thus this difference alone would lead to a maximum velocity about $\sqrt{7} \approx 2.6$ times larger than that of Iceland.

The alternative assumption is the active upwelling is relatively large, perhaps about 10 times larger than the passive upwelling \citep{Ito1999,Maclennan2001}, albeit this ratio decreases away from the bottom of the melting region.
Accounting for a large active upwelling would lead to a maximum velocity about $\sqrt{10} \approx 3.2$ times smaller than that of Iceland at a similarly slow-spreading ridge. 
However, at a fast-spreading ridge, the maximum velocity would only be about about $\sqrt{10/7} \approx 1.2$ times smaller than that of Iceland.

Second, the plume influence on Iceland potentially increases the maximum degree of melting there due to higher potential temperature \citep{Dalton2014} and anomalous source composition \citep{Shorttle2011}.
The exact magnitude of this effect is somewhat uncertain, but simplified petrological models suggest that it might be around 1.5--2 times the normal amount \citep{mckenzie84}. 
The elevated degree of melting is consistent with the enhanced crustal thickness in Iceland of about 20~km, with evidence of a deeper second layer in some regions \citep{jenkins18}.
At the upper end of the range, the effect on maximum melt velocity would be to reduce it by a factor of about $1/\sqrt{2} \approx 0.7$.

Thus at a slow-spreading ridge (i.e. the same spreading rate as Iceland but half the maximum degree of melting), our preferred estimate of the steady-state maximum velocity is 40~m/yr. 
For the East-Pacific-Rise, a fast-spreading ridge, our preferred estimate of maximum velocity is 110~m/yr and a lower bound is 55~m/yr. 
Under the hypothesis that active upwelling beneath Iceland is 10 times faster than passive upwelling, our preferred estimate for the East-Pacific-Rise of maximum velocity is 35~m/yr and a lower bound is 18~m/yr.
Even at the lower end, the travel time across the melt region would be only about 7~kyr.

\subsection{Limitations of analysis} \label{sec:limitations}
In this paper, we have focussed on modelling one aspect of the magmatic response to deglaciation carefully, namely the nonlinear far from steady-state nature of the system.
In doing so we neglected many other complexities, some of which have been considered by previous studies, others of which remain as questions for future research. 

One major simplification of our study is being one dimensional. 
\citet{Armitage2019} and \citet{Eksinchol2019} used quasi-two and three-dimensional models (respectively), which can account for the geographical variation in magmatism and chemical transport.
In particular, \citet{Eksinchol2019} used an axisymmetric ice sheet and a linear ridge and found that the decompression rate varied strongly in space, whereas here we assumed it was a constant.
However, even these quasi-higher-dimensional models have a fundamentally one-dimensional view of melt transport.
Crucially, such models must make some assumption about how melt travels after it reaches the top of the melting region. 
One `end-member' assumption, used by \citet{Eksinchol2019}, is that melts continue to rise at the same rate that they would have done had they remained in the melting region. 
The opposite extreme is that melts would travel extremely rapidly along a decompaction channel at the top of the melting region \citep{sparks91}, leading to instantaneous focussing.
This assumption was used by \citet{burley2015}, for example.
In \ref{app:geometry}, we show that a one-dimensional geometry is a middle ground between these two-dimensional models, and so is arguably a reasonable choice.
\citet{Armitage2019}  argued that the Niobium record gives some evidence for the usefulness of one-dimensional models in capturing the overall fluxes over their quasi-two-dimensional model.
Potentially, the study of \citet{Eksinchol2019} should be revisited to account the possibility of rapid melt focussing, as there might be some signature of this in the Lanthanum data they consider. 

However, while it seems not unreasonable to use a one-dimensional model in preference to quasi-two-dimensional models, other higher dimensional effects are possible. 
Geological evidence from dunite channels in ophiolites, laboratory experiments and numerical modelling indicate that melt transport can be heterogenous due to flow channelization \citep{kelemen95,kelemen00,liang10,pec15,pec17,keller2017,reesjones2018-jfm}.
This can potentially occur through a substantial depth of the melting region up to about 80~km \citep{keller2017}.
It would be worthwhile to use time-dependent two-dimensional numerical models to assess the potential role of flow channelization during deglaciation.

A second potentially important issue is the crustal magmatic system, which has been widely neglected in modelling studies.
Indeed, the idea that deglaciation triggered the release of magma from crustal storage has been discussed as an alternative explanation to fast melt transport \citep{kelemen97}.
There is some major element geochemical evidence for this effect \citep{Gee1998-JPet}.
However, trace element geochemistry suggests that a significant part of the extra emissions following deglaciation come from changes to mantle melting \citep{maclennan02}, which motivated our approach.
However, it may be worth revisiting this question in future.

Other simplifications that we made are likely to be smaller and have all been considered by previous studies. 
These include more complex melting models (such as the role of volatiles such as carbon), the elastic response of the system (which means that the change in melting rate decays with depth) and postglacial rebound.
The nonlinear correction factor we identify, being a ratio of velocities from nonlinear and linear models both making the same set of simplifications, is not likely to be strongly affected.

\section{Conclusions} \label{sec:conclusions}
This study broadly supports the idea that melt velocities in the mantle beneath mid-ocean ridges are relatively fast, at least an order of magnitude greater than the 1~m/yr inferred from microstructural models and measurements assuming diffuse porous flow. 
Melt velocities beneath Iceland during its deglaciation were somewhat higher than usual. 
In quantitative terms, we found that even though the melting rate was perhaps 30 times greater than usual, the effective velocity was only about 3 times faster.
So the implication of fast melt velocities beneath Iceland drawn by previous studies stands, although the best estimate of the steady-state maximum melt velocity is reduced from 100~m/yr to 30~m/yr.
The plume influence on Iceland leads to a higher degree of melting and could potentially cause faster active mantle upwelling.
Accounting for the higher degree of melting would reduce melt velocities at other ridges by a factor of no more than about 0.7. 
However, this effect is generally more than compensated by the faster spreading rate at most other ridges relative to Iceland.
Faster active mantle upwelling combined with the higher degree of melting at Iceland would reduce melt velocities at the East-Pacific-Rise, for example, by a factor of about 0.6. 
So fast maximum melt velocities greater than 10~m/yr, and hence low residual porosities, are probably a general feature of mid-ocean ridges, consistent with Uranium-series and geophysical estimates.  
This estimate of maximum melt velocity is an important constraint on the results of \citet{Cerpa2019} and suggests that glacial/interglacial sea-level variation in the Pleistocene caused variation in mid-ocean ridge carbon fluxes with a lag of less than 2~kyr. 
The relative shortness of this lag appears to be inconsistent with the suggestion by \citet{huybers2017} that delayed variation in mid-ocean carbon fluxes paced 100~kyr Pleistocene climate cycles.
More broadly, our estimates of melt velocity from Iceland can be used to infer melt velocity for any particular ridge segment of interest using the methodology we developed. 
Such estimates can then be used when interpreting geochemical measurements from mid-ocean ridge basalts there. 

\section*{Acknowledgements}
Code to reproduce the calculations made and figures plotted is available \citep{ReesJones19Code}.
We thank the Leverhulme Trust for financial support.
We thank J. Maclennan and R. Katz for helpful discussions about an earlier version of this manuscript.

\appendix
\section{Formulae for additional emissions at the top of the melting column} \label{app:solution}  
There are two main cases which are separated by the special case in which a parcel of melt travels from the bottom to the top of the melting column over precisely the deglaciation period.
We first introduce some additional notation defining two time scales that separate various regimes:
\begin{eqnarray}
 \hat{t}_1 & = & \left(1+\mathcal{A} \right)^{-1/2}, \\
  \hat{t}_2 & = & \left[ 1 + \mathcal{A}\left(1+\mathcal{A} \right)\lambda^2 \right]^{1/2} - \mathcal{A} \lambda,
\end{eqnarray}
where $\hat{t}_1$ is the travel time for a parcel of extra melt leaving $\hat{z}=0$ at  $\hat{t}=0$ assuming that it reaches the top while the deglaciation event is ongoing and $\hat{t}_2$ is the travel time if it reaches the top after deglaciation finishes (see Supplementary Material, figure S1).
   
\subsection{Case 1: $\lambda > \hat{t}_1$}
In this case, melt transport is sufficiently rapid that a parcel of extra melt rises from bottom to top before deglaciation finishes. 
We find, using the method of characteristics (see Supplementary Material for a full derivation), that the additional melt flux caused by deglaciation is given by:
\begin{equation} 
    \hat{Q} = \left\{\begin{array}{ll}
         \hat{Q}_1 & (0\leq \hat{t} \leq \hat{t}_1),\\
        1 & (\hat{t}_1\leq \hat{t} \leq \lambda),\\
         \hat{Q}_3 & (\lambda \leq \hat{t} \leq 1+\lambda),
        \end{array}\right.
\end{equation}  
where
\begin{equation}
\hat{Q}_1 = 2 \hat{t} \left(1-t^2\mathcal{A} \right)^{1/2} - \hat{t}^2(1-\mathcal{A}), 
\end{equation}
\begin{equation}
\hat{Q}_3 = 1+ (\hat{t}-\lambda)^2 (1+2\mathcal{A}) - 2(1+\mathcal{A})^{1/2}(\hat{t}-\lambda)\left[1+\mathcal{A}(\hat{t}-\lambda)^2\right]^{1/2} .
\end{equation}
Note the period ($\hat{t}_1\leq \hat{t} \leq \lambda$) during which $\hat{Q}$ is saturated at its maximum value.
All the parcels of melt reaching the top of the column during this period have accumulated the extra melt generated by deglaciation throughout the column.

\subsection{Case 2: $\lambda < \hat{t}_1$}
In this case, melt transport is slow enough that a parcel of extra melt does not rise from bottom to top before deglaciation finishes. 
The additional melt flux caused by deglaciation is given by:
\begin{equation} 
    \hat{Q} = \left\{\begin{array}{ll}
         \hat{Q}_1 & (0\leq \hat{t} \leq \lambda),\\
        \hat{Q}_2 & (\lambda \leq \hat{t} \leq \hat{t}_2),\\
         \hat{Q}_3 & (\hat{t}_2 \leq \hat{t} \leq 1+\lambda),
        \end{array}\right.
\end{equation}  
where
\begin{equation}
 \hat{Q}_2  =  \lambda \left\{-2\hat{t} +\lambda(1+\mathcal{A}) +2 \left[1+\mathcal{A} \lambda (\lambda-2\hat{t} )\right]^{1/2} \right\} .
\end{equation}
Rather than saturating in the second period, the melt flux $\hat{Q}_2$ decreases over time because the deglaciation period has finished.

\subsection{Cumulative emissions} \label{app:cumulative}
The cumulative emissions are obtained by integrating these formulae with respect to time.
In general, the expressions are rather complicated (see Supplementary Material for full details). 
However, one special case is fairly simple and helpful in estimating the correction factors described in section~\ref{sec:NL_correction}.

For sufficiently fast melt transport (Case 1), most of the extra emissions occur during the period when $\hat{Q}=1$, so cumulative emissions rise linearly with time. 
So provided $\lambda \geq \hat{t} \geq \hat{t}_1$, the relative cumulative emissions are given by
\begin{equation}
\hat{F} =\lambda^{-1} \left\{ \hat{t} -  \frac{2}{3\mathcal{A}} \left[ (1+\mathcal{A})^{1/2} - 1\right] \right\}.
\end{equation}
We want to determine the time $\hat{t}_\mathrm{c}$ it takes to achieve a certain fraction of the total emissions $\hat{F}_\mathrm{c}$. 
This satisfies
\begin{equation} \label{eq:t_crit}
\hat{t}_c = \hat{F}_c \lambda + \Delta \hat{t} (\mathcal{A}),
\end{equation}
where $\Delta \hat{t} (\mathcal{A}) = (2/{3\mathcal{A}})  \left[ (1+\mathcal{A})^{1/2} - 1\right]$.
Note that $\Delta \hat{t} (\mathcal{A} \rightarrow 0) = 1/3$ (by L'H\^{o}pital's Rule).

Following the methodology described in section~\ref{sec:NL_correction}, a linear model  ($\mathcal{A} \rightarrow 0$, denoted superscript $^L$) and a nonlinear model  ($\mathcal{A} > 0$, denoted superscript $^{NL}$)  are equivalent when 
\begin{equation}
\frac{\hat{t}_c^{L}}{\lambda^{L}} = \frac{\hat{t}_c^{NL}}{\lambda^{NL}}, 
\end{equation}
in which we used the non-dimensionalization of time.
By combining this expression with equation~(\ref{eq:t_crit}), we find
\begin{equation} 
\frac{\lambda^{NL}}{\lambda^{L}} = 2 \mathcal{A}^{-1} \left[ (1+\mathcal{A})^{1/2} - 1\right], 
\end{equation}
which gives the correction factor for the maximum velocity estimate shown in equation~(\ref{eq:correction_factor}).

\section{Geometrical effects} \label{app:geometry}
The melting region underneath Iceland is not one-dimensional.
Here we develop simple geometrical arguments to explore some possible two-dimensional effects (see the main text, section~\ref{sec:limitations}, for discussion of two-dimensional effects beyond the geometrical).

\begin{figure}
\noindent\includegraphics[width=0.85\linewidth]{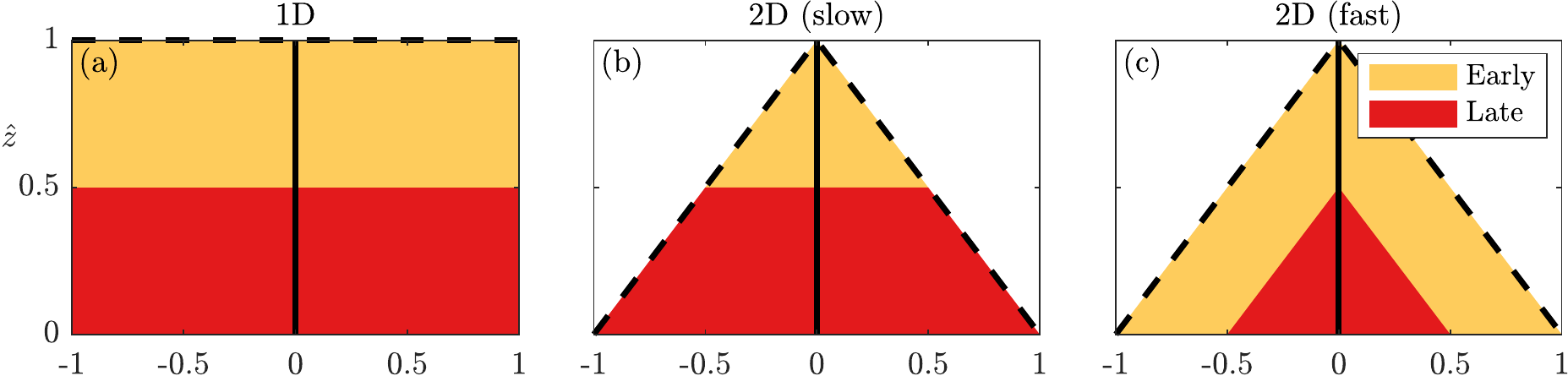}
\caption{Geometrical effects. The melting region is coloured yellow for melts that arrive early and red for melts that arrive late. Solid black lines denote the ridge axis and dashed black lines denote the top of the melting region. (a) One-dimensional melting region. (b) Two-dimensional melting region in which travel time does not depend on horizontal distance to the ridge (labelled `slow'). (c) Two-dimensional melting region in which travel time depends on distance to the top of the melting region (labelled `fast').  }
\label{fig:geometry}
\end{figure}  

Figure~\ref{fig:geometry} shows a melting region divided up according to the starting depth for melts that erupt either `early' and `late' (say before and after deglaciation finishes). 
For a one-dimensional geometry, these periods correspond to a constant height $\hat{z}_0$.
The example plotted shows $\hat{z}_0=0.5$.
For a possible two-dimensional triangular geometry, the division depends on assumptions about melt transport. 
In particular, assuming a constant height $\hat{z}_0$ is equivalent to assuming one-dimensional melt transport continuing at the same rate it would have done had it not encountered the top of the melting region.
An alternative possibility is assuming a constant distance from the top of the melting region. 
For the simplest possible case of uniform melt transport (a special case of our model when $n=1$ and $\mathcal{A} \ll 1$), this is exactly equivalent to a uniform travel time and very rapid melt transport from the top of the melting region to the surface. 
For this reason, we label the uniform distance case `fast' and the uniform height case `slow.'
Of course, more complex criteria could be used, but these examples are particularly simple and bracket a plausible range of behaviour in terms of melt transport from the top of the melting region. 

An important way to characterize the effect of geometry is to define the ratio $R$ of the area of the early region to the area of the late region. 
Note that this ratio is the same for any triangular region, it does not depend on the angle of the melting region. 
These are calculated as:
\begin{equation} 
     R = \left\{\begin{array}{ll}
        1-\hat{z}_0 & (\mathrm{1D}),\\
         (1-\hat{z}_0)^2 & (\mathrm{2D, slow}),\\
        1-\hat{z}_0^2 & (\mathrm{2D, fast}).
        \end{array}\right.
\end{equation} 
An important observation is that the ratio for a one-dimensional geometry is the arithmetic mean of the pair of two-dimensional geometries (regardless of $\hat{z}_0$). 
Indeed, this result is also clear geometrically (figure~\ref{fig:geometry}).

Precise quantitative details will of course depend on the definitions of early and late regions.
The decrease in the mantle upwelling rate at the base of the column and height of the melting region away from the ridge axis will also affect quantitative results.
Nevertheless, this simple example illustrates the crucial qualitative idea: late melts are more significant for a triangular melting region provided melt transport (focussing) from the top of the region is relatively slow, but less significant if focussing is rapid. 
Therefore, a one-dimensional model is not unreasonable, since it is an average between these extremes.

\bibliography{deglaciation}

\begin{thebibliography}{45}%
\makeatletter
\providecommand \@ifxundefined [1]{%
 \@ifx{#1\undefined}
}%
\providecommand \@ifnum [1]{%
 \ifnum #1\expandafter \@firstoftwo
 \else \expandafter \@secondoftwo
 \fi
}%
\providecommand \@ifx [1]{%
 \ifx #1\expandafter \@firstoftwo
 \else \expandafter \@secondoftwo
 \fi
}%
\providecommand \natexlab [1]{#1}%
\providecommand \enquote  [1]{``#1''}%
\providecommand \bibnamefont  [1]{#1}%
\providecommand \bibfnamefont [1]{#1}%
\providecommand \citenamefont [1]{#1}%
\providecommand \href@noop [0]{\@secondoftwo}%
\providecommand \href [0]{\begingroup \@sanitize@url \@href}%
\providecommand \@href[1]{\@@startlink{#1}\@@href}%
\providecommand \@@href[1]{\endgroup#1\@@endlink}%
\providecommand \@sanitize@url [0]{\catcode `\\12\catcode `\$12\catcode
  `\&12\catcode `\#12\catcode `\^12\catcode `\_12\catcode `\%12\relax}%
\providecommand \@@startlink[1]{}%
\providecommand \@@endlink[0]{}%
\providecommand \url  [0]{\begingroup\@sanitize@url \@url }%
\providecommand \@url [1]{\endgroup\@href {#1}{\urlprefix }}%
\providecommand \urlprefix  [0]{URL }%
\providecommand \Eprint [0]{\href }%
\providecommand \doibase [0]{http://dx.doi.org/}%
\providecommand \selectlanguage [0]{\@gobble}%
\providecommand \bibinfo  [0]{\@secondoftwo}%
\providecommand \bibfield  [0]{\@secondoftwo}%
\providecommand \translation [1]{[#1]}%
\providecommand \BibitemOpen [0]{}%
\providecommand \bibitemStop [0]{}%
\providecommand \bibitemNoStop [0]{.\EOS\space}%
\providecommand \EOS [0]{\spacefactor3000\relax}%
\providecommand \BibitemShut  [1]{\csname bibitem#1\endcsname}%
\let\auto@bib@innerbib\@empty
\bibitem [{\citenamefont {Connolly}\ \emph {et~al.}(2009)\citenamefont
  {Connolly}, \citenamefont {Schmidt}, \citenamefont {Solferino},\ and\
  \citenamefont {Bagdassarov}}]{connolly09}%
  \BibitemOpen
  \bibfield  {author} {\bibinfo {author} {\bibfnamefont {J.~A.~D.}\
  \bibnamefont {Connolly}}, \bibinfo {author} {\bibfnamefont {M.~W.}\
  \bibnamefont {Schmidt}}, \bibinfo {author} {\bibfnamefont {G.}~\bibnamefont
  {Solferino}}, \ and\ \bibinfo {author} {\bibfnamefont {N.}~\bibnamefont
  {Bagdassarov}},\ }\href {\doibase 10.1038/nature08517} {\bibfield  {journal}
  {\bibinfo  {journal} {Nature}\ }\textbf {\bibinfo {volume} {462}},\ \bibinfo
  {pages} {209} (\bibinfo {year} {2009})}\BibitemShut {NoStop}%
\bibitem [{\citenamefont {Wark}\ \emph {et~al.}(2003)\citenamefont {Wark},
  \citenamefont {Williams}, \citenamefont {Watson},\ and\ \citenamefont
  {Price}}]{wark03}%
  \BibitemOpen
  \bibfield  {author} {\bibinfo {author} {\bibfnamefont {D.}~\bibnamefont
  {Wark}}, \bibinfo {author} {\bibfnamefont {C.}~\bibnamefont {Williams}},
  \bibinfo {author} {\bibfnamefont {E.}~\bibnamefont {Watson}}, \ and\ \bibinfo
  {author} {\bibfnamefont {J.}~\bibnamefont {Price}},\ }\href {\doibase
  10.1029/2001JB001575} {\bibfield  {journal} {\bibinfo  {journal} {J.\
  Geophys.\ Res.}\ }\textbf {\bibinfo {volume} {108}} (\bibinfo {year}
  {2003}),\ 10.1029/2001JB001575}\BibitemShut {NoStop}%
\bibitem [{\citenamefont {Zhu}\ and\ \citenamefont {Hirth}(2003)}]{zhu03}%
  \BibitemOpen
  \bibfield  {author} {\bibinfo {author} {\bibfnamefont {W.}~\bibnamefont
  {Zhu}}\ and\ \bibinfo {author} {\bibfnamefont {G.}~\bibnamefont {Hirth}},\
  }\href {\doibase 10.1016/S0012-821X(03)00264-4} {\bibfield  {journal}
  {\bibinfo  {journal} {Earth. Planet. Sci. Lett.}\ }\textbf {\bibinfo {volume}
  {212}},\ \bibinfo {pages} {407} (\bibinfo {year} {2003})}\BibitemShut
  {NoStop}%
\bibitem [{\citenamefont {{Rudge}}(2018)}]{Rudge18}%
  \BibitemOpen
  \bibfield  {author} {\bibinfo {author} {\bibfnamefont {J.~F.}\ \bibnamefont
  {{Rudge}}},\ }\href {\doibase 10.1098/rspa.2017.0639} {\bibfield  {journal}
  {\bibinfo  {journal} {Proc.\ Roy.\ Soc.\ A}\ }\textbf {\bibinfo {volume}
  {474}} (\bibinfo {year} {2018}),\ 10.1098/rspa.2017.0639}\BibitemShut
  {NoStop}%
\bibitem [{\citenamefont {Stracke}\ \emph {et~al.}(2006)\citenamefont
  {Stracke}, \citenamefont {Bourdon},\ and\ \citenamefont
  {McKenzie}}]{stracke06}%
  \BibitemOpen
  \bibfield  {author} {\bibinfo {author} {\bibfnamefont {A.}~\bibnamefont
  {Stracke}}, \bibinfo {author} {\bibfnamefont {B.}~\bibnamefont {Bourdon}}, \
  and\ \bibinfo {author} {\bibfnamefont {D.}~\bibnamefont {McKenzie}},\ }\href
  {\doibase 10.1016/j.epsl.2006.01.057} {\bibfield  {journal} {\bibinfo
  {journal} {Earth Plan.\ Sci.\ Lett.}\ }\textbf {\bibinfo {volume} {244}},\
  \bibinfo {pages} {97} (\bibinfo {year} {2006})}\BibitemShut {NoStop}%
\bibitem [{\citenamefont {Elliott}\ and\ \citenamefont
  {Spiegelman}(2014)}]{Elliot2014}%
  \BibitemOpen
  \bibfield  {author} {\bibinfo {author} {\bibfnamefont {T.}~\bibnamefont
  {Elliott}}\ and\ \bibinfo {author} {\bibfnamefont {M.}~\bibnamefont
  {Spiegelman}},\ }in\ \href {\doibase 10.1016/B978-0-08-095975-7.00317-X}
  {\emph {\bibinfo {booktitle} {Treatise on {G}eochemistry}}},\ \bibinfo
  {editor} {edited by\ \bibinfo {editor} {\bibfnamefont {H.~D.}\ \bibnamefont
  {Holland}}\ and\ \bibinfo {editor} {\bibfnamefont {K.~K.}\ \bibnamefont
  {Turekian}}}\ (\bibinfo  {publisher} {Elsevier},\ \bibinfo {year} {2014})\
  \bibinfo {edition} {2nd}\ ed.,\ pp.\ \bibinfo {pages} {543--581}\BibitemShut
  {NoStop}%
\bibitem [{\citenamefont {Turner}\ \emph {et~al.}(2001)\citenamefont {Turner},
  \citenamefont {Evans},\ and\ \citenamefont {Hawkesworth}}]{Turner2001}%
  \BibitemOpen
  \bibfield  {author} {\bibinfo {author} {\bibfnamefont {S.}~\bibnamefont
  {Turner}}, \bibinfo {author} {\bibfnamefont {P.}~\bibnamefont {Evans}}, \
  and\ \bibinfo {author} {\bibfnamefont {C.}~\bibnamefont {Hawkesworth}},\
  }\href {\doibase 10.1126/science.1059904} {\bibfield  {journal} {\bibinfo
  {journal} {Science}\ }\textbf {\bibinfo {volume} {292}},\ \bibinfo {pages}
  {1363} (\bibinfo {year} {2001})}\BibitemShut {NoStop}%
\bibitem [{\citenamefont {Rubin}\ \emph {et~al.}(2005)\citenamefont {Rubin},
  \citenamefont {van~der Zander}, \citenamefont {Smith},\ and\ \citenamefont
  {Bergmanis}}]{rubin05}%
  \BibitemOpen
  \bibfield  {author} {\bibinfo {author} {\bibfnamefont {K.~H.}\ \bibnamefont
  {Rubin}}, \bibinfo {author} {\bibfnamefont {I.}~\bibnamefont {van~der
  Zander}}, \bibinfo {author} {\bibfnamefont {M.~C.}\ \bibnamefont {Smith}}, \
  and\ \bibinfo {author} {\bibfnamefont {E.~C.}\ \bibnamefont {Bergmanis}},\
  }\href {\doibase 10.1038/nature03993} {\bibfield  {journal} {\bibinfo
  {journal} {Nature}\ }\textbf {\bibinfo {volume} {437}},\ \bibinfo {pages}
  {534} (\bibinfo {year} {2005})}\BibitemShut {NoStop}%
\bibitem [{\citenamefont {Turner}\ \emph {et~al.}(2012)\citenamefont {Turner},
  \citenamefont {Reagan}, \citenamefont {Vigier},\ and\ \citenamefont
  {Bourdon}}]{Turner2012}%
  \BibitemOpen
  \bibfield  {author} {\bibinfo {author} {\bibfnamefont {S.}~\bibnamefont
  {Turner}}, \bibinfo {author} {\bibfnamefont {M.}~\bibnamefont {Reagan}},
  \bibinfo {author} {\bibfnamefont {N.}~\bibnamefont {Vigier}}, \ and\ \bibinfo
  {author} {\bibfnamefont {B.}~\bibnamefont {Bourdon}},\ }\href {\doibase
  10.1029/2012GC004173} {\bibfield  {journal} {\bibinfo  {journal} {Geochem.\
  Geophys.\ Geosys.}\ }\textbf {\bibinfo {volume} {13}} (\bibinfo {year}
  {2012}),\ 10.1029/2012GC004173}\BibitemShut {NoStop}%
\bibitem [{\citenamefont {Armitage}\ \emph {et~al.}(2019)\citenamefont
  {Armitage}, \citenamefont {Ferguson}, \citenamefont {Petersen},\ and\
  \citenamefont {Creyts}}]{Armitage2019}%
  \BibitemOpen
  \bibfield  {author} {\bibinfo {author} {\bibfnamefont {J.~J.}\ \bibnamefont
  {Armitage}}, \bibinfo {author} {\bibfnamefont {D.~J.}\ \bibnamefont
  {Ferguson}}, \bibinfo {author} {\bibfnamefont {K.~D.}\ \bibnamefont
  {Petersen}}, \ and\ \bibinfo {author} {\bibfnamefont {T.~T.}\ \bibnamefont
  {Creyts}},\ }\href {\doibase 10.1029/2019GL081955} {\bibfield  {journal}
  {\bibinfo  {journal} {Geophys. Res. Letts.}\ }\textbf {\bibinfo {volume}
  {46}},\ \bibinfo {pages} {6451} (\bibinfo {year} {2019})}\BibitemShut
  {NoStop}%
\bibitem [{\citenamefont {Eksinchol}\ \emph {et~al.}(2019)\citenamefont
  {Eksinchol}, \citenamefont {Rudge},\ and\ \citenamefont
  {Maclennan}}]{Eksinchol2019}%
  \BibitemOpen
  \bibfield  {author} {\bibinfo {author} {\bibfnamefont {I.}~\bibnamefont
  {Eksinchol}}, \bibinfo {author} {\bibfnamefont {J.~F.}\ \bibnamefont
  {Rudge}}, \ and\ \bibinfo {author} {\bibfnamefont {J.}~\bibnamefont
  {Maclennan}},\ }\href {\doibase 10.1029/2019GC008222} {\bibfield  {journal}
  {\bibinfo  {journal} {Geochem.\ Geophys.\ Geosys.}\ }\textbf {\bibinfo
  {volume} {20}} (\bibinfo {year} {2019}),\ 10.1029/2019GC008222}\BibitemShut
  {NoStop}%
\bibitem [{\citenamefont {Jull}\ and\ \citenamefont {McKenzie}(1996)}]{jull96}%
  \BibitemOpen
  \bibfield  {author} {\bibinfo {author} {\bibfnamefont {M.}~\bibnamefont
  {Jull}}\ and\ \bibinfo {author} {\bibfnamefont {D.}~\bibnamefont
  {McKenzie}},\ }\href {\doibase 10.1029/96JB01308} {\bibfield  {journal}
  {\bibinfo  {journal} {J.\ Geophys.\ Res.}\ }\textbf {\bibinfo {volume}
  {101}},\ \bibinfo {pages} {21815} (\bibinfo {year} {1996})}\BibitemShut
  {NoStop}%
\bibitem [{\citenamefont {Maclennan}\ \emph {et~al.}(2002)\citenamefont
  {Maclennan}, \citenamefont {Jull}, \citenamefont {McKenzie}, \citenamefont
  {Slater},\ and\ \citenamefont {Gr\"{o}nvold}}]{maclennan02}%
  \BibitemOpen
  \bibfield  {author} {\bibinfo {author} {\bibfnamefont {J.}~\bibnamefont
  {Maclennan}}, \bibinfo {author} {\bibfnamefont {M.}~\bibnamefont {Jull}},
  \bibinfo {author} {\bibfnamefont {D.}~\bibnamefont {McKenzie}}, \bibinfo
  {author} {\bibfnamefont {L.}~\bibnamefont {Slater}}, \ and\ \bibinfo {author}
  {\bibfnamefont {K.}~\bibnamefont {Gr\"{o}nvold}},\ }\href {\doibase
  10.1029/2001GC000282} {\bibfield  {journal} {\bibinfo  {journal} {Geochem.\
  Geophys.\ Geosys.}\ } (\bibinfo {year} {2002}),\
  10.1029/2001GC000282}\BibitemShut {NoStop}%
\bibitem [{\citenamefont {Schmidt}\ \emph {et~al.}(2013)\citenamefont
  {Schmidt}, \citenamefont {Lund}, \citenamefont {Hieronymus}, \citenamefont
  {Maclennan}, \citenamefont {\'{A}rnad\'{o}ttir},\ and\ \citenamefont
  {Pagli}}]{Schmidt2013}%
  \BibitemOpen
  \bibfield  {author} {\bibinfo {author} {\bibfnamefont {P.}~\bibnamefont
  {Schmidt}}, \bibinfo {author} {\bibfnamefont {B.}~\bibnamefont {Lund}},
  \bibinfo {author} {\bibfnamefont {C.}~\bibnamefont {Hieronymus}}, \bibinfo
  {author} {\bibfnamefont {J.}~\bibnamefont {Maclennan}}, \bibinfo {author}
  {\bibfnamefont {T.}~\bibnamefont {\'{A}rnad\'{o}ttir}}, \ and\ \bibinfo
  {author} {\bibfnamefont {C.}~\bibnamefont {Pagli}},\ }\href {\doibase
  10.1002/jgrb.50273} {\bibfield  {journal} {\bibinfo  {journal} {J. Geophys.
  Res. -- Solid Earth}\ }\textbf {\bibinfo {volume} {118}},\ \bibinfo {pages}
  {3366} (\bibinfo {year} {2013})}\BibitemShut {NoStop}%
\bibitem [{\citenamefont {Tuffen}(2010)}]{Tuffen2010}%
  \BibitemOpen
  \bibfield  {author} {\bibinfo {author} {\bibfnamefont {H.}~\bibnamefont
  {Tuffen}},\ }\href {\doibase 10.1098/rsta.2010.0063} {\bibfield  {journal}
  {\bibinfo  {journal} {Philos. Trans. R. Soc. A}\ }\textbf {\bibinfo {volume}
  {368}},\ \bibinfo {pages} {2535} (\bibinfo {year} {2010})}\BibitemShut
  {NoStop}%
\bibitem [{\citenamefont {Lund}\ and\ \citenamefont {Asimow}(2011)}]{lund2011}%
  \BibitemOpen
  \bibfield  {author} {\bibinfo {author} {\bibfnamefont {D.~C.}\ \bibnamefont
  {Lund}}\ and\ \bibinfo {author} {\bibfnamefont {P.~D.}\ \bibnamefont
  {Asimow}},\ }\href {\doibase 10.1029/2011GC003693} {\bibfield  {journal}
  {\bibinfo  {journal} {Geochem.\ Geophys.\ Geosys.}\ }\textbf {\bibinfo
  {volume} {12}} (\bibinfo {year} {2011}),\ 10.1029/2011GC003693}\BibitemShut
  {NoStop}%
\bibitem [{\citenamefont {Miller}\ \emph {et~al.}(2014)\citenamefont {Miller},
  \citenamefont {Zhu}, \citenamefont {Mont{\'e}si},\ and\ \citenamefont
  {Gaetani}}]{miller14}%
  \BibitemOpen
  \bibfield  {author} {\bibinfo {author} {\bibfnamefont {K.~J.}\ \bibnamefont
  {Miller}}, \bibinfo {author} {\bibfnamefont {W.}~\bibnamefont {Zhu}},
  \bibinfo {author} {\bibfnamefont {L.~G.~J.}\ \bibnamefont {Mont{\'e}si}}, \
  and\ \bibinfo {author} {\bibfnamefont {G.~A.}\ \bibnamefont {Gaetani}},\
  }\href {\doibase 10.1016/j.epsl.2013.12.003} {\bibfield  {journal} {\bibinfo
  {journal} {Earth Plan.\ Sci.\ Lett.}\ }\textbf {\bibinfo {volume} {388}},\
  \bibinfo {pages} {273} (\bibinfo {year} {2014})}\BibitemShut {NoStop}%
\bibitem [{\citenamefont {Shorttle}\ and\ \citenamefont
  {Maclennan}(2011)}]{Shorttle2011}%
  \BibitemOpen
  \bibfield  {author} {\bibinfo {author} {\bibfnamefont {O.}~\bibnamefont
  {Shorttle}}\ and\ \bibinfo {author} {\bibfnamefont {J.}~\bibnamefont
  {Maclennan}},\ }\href {\doibase 10.1029/2011GC003748} {\bibfield  {journal}
  {\bibinfo  {journal} {Geochem.\ Geophys.\ Geosys.}\ }\textbf {\bibinfo
  {volume} {12}} (\bibinfo {year} {2011}),\ 10.1029/2011GC003748}\BibitemShut
  {NoStop}%
\bibitem [{\citenamefont {Huybers}\ and\ \citenamefont
  {Langmuir}(2009)}]{huybers2009}%
  \BibitemOpen
  \bibfield  {author} {\bibinfo {author} {\bibfnamefont {P.}~\bibnamefont
  {Huybers}}\ and\ \bibinfo {author} {\bibfnamefont {C.}~\bibnamefont
  {Langmuir}},\ }\href {\doibase 10.1016/j.epsl.2009.07.014} {\bibfield
  {journal} {\bibinfo  {journal} {Earth Planet. Sci. Lett.}\ }\textbf {\bibinfo
  {volume} {286}},\ \bibinfo {pages} {479} (\bibinfo {year}
  {2009})}\BibitemShut {NoStop}%
\bibitem [{\citenamefont {Crowley}\ \emph {et~al.}(2015)\citenamefont
  {Crowley}, \citenamefont {Katz}, \citenamefont {Huybers}, \citenamefont
  {Langmuir},\ and\ \citenamefont {Park}}]{crowley2015}%
  \BibitemOpen
  \bibfield  {author} {\bibinfo {author} {\bibfnamefont {J.~W.}\ \bibnamefont
  {Crowley}}, \bibinfo {author} {\bibfnamefont {R.~F.}\ \bibnamefont {Katz}},
  \bibinfo {author} {\bibfnamefont {P.}~\bibnamefont {Huybers}}, \bibinfo
  {author} {\bibfnamefont {C.~H.}\ \bibnamefont {Langmuir}}, \ and\ \bibinfo
  {author} {\bibfnamefont {S.-H.}\ \bibnamefont {Park}},\ }\href {\doibase
  10.1126/science.1261508} {\bibfield  {journal} {\bibinfo  {journal}
  {Science}\ }\textbf {\bibinfo {volume} {347}},\ \bibinfo {pages} {1237}
  (\bibinfo {year} {2015})}\BibitemShut {NoStop}%
\bibitem [{\citenamefont {Burley}\ and\ \citenamefont
  {Katz}(2015)}]{burley2015}%
  \BibitemOpen
  \bibfield  {author} {\bibinfo {author} {\bibfnamefont {J.~M.}\ \bibnamefont
  {Burley}}\ and\ \bibinfo {author} {\bibfnamefont {R.~F.}\ \bibnamefont
  {Katz}},\ }\href {\doibase 10.1016/j.epsl.2015.06.031} {\bibfield  {journal}
  {\bibinfo  {journal} {Earth Planet. Sci. Lett.}\ }\textbf {\bibinfo {volume}
  {426}},\ \bibinfo {pages} {246} (\bibinfo {year} {2015})}\BibitemShut
  {NoStop}%
\bibitem [{\citenamefont {Huybers}\ and\ \citenamefont
  {Langmuir}(2017)}]{huybers2017}%
  \BibitemOpen
  \bibfield  {author} {\bibinfo {author} {\bibfnamefont {P.}~\bibnamefont
  {Huybers}}\ and\ \bibinfo {author} {\bibfnamefont {C.~H.}\ \bibnamefont
  {Langmuir}},\ }\href {\doibase 10.1016/j.epsl.2016.09.021} {\bibfield
  {journal} {\bibinfo  {journal} {Earth Planet. Sci. Lett.}\ }\textbf {\bibinfo
  {volume} {457}},\ \bibinfo {pages} {238} (\bibinfo {year}
  {2017})}\BibitemShut {NoStop}%
\bibitem [{\citenamefont {Cerpa}\ \emph {et~al.}(2019)\citenamefont {Cerpa},
  \citenamefont {Rees~Jones},\ and\ \citenamefont {Katz}}]{Cerpa2019}%
  \BibitemOpen
  \bibfield  {author} {\bibinfo {author} {\bibfnamefont {N.~G.}\ \bibnamefont
  {Cerpa}}, \bibinfo {author} {\bibfnamefont {D.~W.}\ \bibnamefont
  {Rees~Jones}}, \ and\ \bibinfo {author} {\bibfnamefont {R.~F.}\ \bibnamefont
  {Katz}},\ }\href {\doibase 10.1016/j.epsl.2019.115845} {\bibfield  {journal}
  {\bibinfo  {journal} {Earth Planet. Sci. Lett.}\ }\textbf {\bibinfo {volume}
  {528}},\ \bibinfo {pages} {115845} (\bibinfo {year} {2019})}\BibitemShut
  {NoStop}%
\bibitem [{\citenamefont {Siddall}\ \emph {et~al.}(2010)\citenamefont
  {Siddall}, \citenamefont {H{\"o}nisch}, \citenamefont {Waelbroeck},\ and\
  \citenamefont {Huybers}}]{siddall2010}%
  \BibitemOpen
  \bibfield  {author} {\bibinfo {author} {\bibfnamefont {M.}~\bibnamefont
  {Siddall}}, \bibinfo {author} {\bibfnamefont {B.}~\bibnamefont
  {H{\"o}nisch}}, \bibinfo {author} {\bibfnamefont {C.}~\bibnamefont
  {Waelbroeck}}, \ and\ \bibinfo {author} {\bibfnamefont {P.}~\bibnamefont
  {Huybers}},\ }\href {\doibase 10.1016/j.quascirev.2009.05.011} {\bibfield
  {journal} {\bibinfo  {journal} {Quat. Sci. Rev.}\ }\textbf {\bibinfo {volume}
  {29}},\ \bibinfo {pages} {170} (\bibinfo {year} {2010})}\BibitemShut
  {NoStop}%
\bibitem [{\citenamefont {McKenzie}(1984)}]{mckenzie84}%
  \BibitemOpen
  \bibfield  {author} {\bibinfo {author} {\bibfnamefont {D.}~\bibnamefont
  {McKenzie}},\ }\href {\doibase 10.1093/petrology/25.3.713} {\bibfield
  {journal} {\bibinfo  {journal} {J.\ Petrol.}\ }\textbf {\bibinfo {volume}
  {25}},\ \bibinfo {pages} {713} (\bibinfo {year} {1984})}\BibitemShut
  {NoStop}%
\bibitem [{\citenamefont {Ribe}(1985)}]{Ribe1985}%
  \BibitemOpen
  \bibfield  {author} {\bibinfo {author} {\bibfnamefont {N.~M.}\ \bibnamefont
  {Ribe}},\ }\href {\doibase 10.1016/0012-821X(85)90084-6} {\bibfield
  {journal} {\bibinfo  {journal} {Earth Planet. Sci. Lett.}\ }\textbf {\bibinfo
  {volume} {73}},\ \bibinfo {pages} {361} (\bibinfo {year} {1985})}\BibitemShut
  {NoStop}%
\bibitem [{\citenamefont {Hewitt}(2010)}]{hewitt10}%
  \BibitemOpen
  \bibfield  {author} {\bibinfo {author} {\bibfnamefont {I.~J.}\ \bibnamefont
  {Hewitt}},\ }\href {\doibase 10.1016/j.epsl.2010.10.010} {\bibfield
  {journal} {\bibinfo  {journal} {Earth Plan. Sci. Lett.}\ }\textbf {\bibinfo
  {volume} {300}},\ \bibinfo {pages} {264} (\bibinfo {year}
  {2010})}\BibitemShut {NoStop}%
\bibitem [{\citenamefont {von Bargen}\ and\ \citenamefont
  {Waff}(1986)}]{vonbargen86}%
  \BibitemOpen
  \bibfield  {author} {\bibinfo {author} {\bibfnamefont {N.}~\bibnamefont {von
  Bargen}}\ and\ \bibinfo {author} {\bibfnamefont {H.~S.}\ \bibnamefont
  {Waff}},\ }\href {\doibase 10.1029/JB091iB09p09261} {\bibfield  {journal}
  {\bibinfo  {journal} {J. Geophys. Res.}\ }\textbf {\bibinfo {volume} {91}},\
  \bibinfo {pages} {9261} (\bibinfo {year} {1986})}\BibitemShut {NoStop}%
\bibitem [{\citenamefont {Scott}\ and\ \citenamefont
  {Stevenson}(1984)}]{scott84}%
  \BibitemOpen
  \bibfield  {author} {\bibinfo {author} {\bibfnamefont {D.~R.}\ \bibnamefont
  {Scott}}\ and\ \bibinfo {author} {\bibfnamefont {D.~J.}\ \bibnamefont
  {Stevenson}},\ }\href {\doibase 10.1029/GL011i011p01161} {\bibfield
  {journal} {\bibinfo  {journal} {Geophys.\ Res.\ Lett.}\ }\textbf {\bibinfo
  {volume} {11}},\ \bibinfo {pages} {1161} (\bibinfo {year}
  {1984})}\BibitemShut {NoStop}%
\bibitem [{\citenamefont {Spiegelman}(1993)}]{spiegelman93a}%
  \BibitemOpen
  \bibfield  {author} {\bibinfo {author} {\bibfnamefont {M.}~\bibnamefont
  {Spiegelman}},\ }\href {\doibase 10.1017/S0022112093000369} {\bibfield
  {journal} {\bibinfo  {journal} {J.\ Fluid Mech.}\ }\textbf {\bibinfo {volume}
  {247}},\ \bibinfo {pages} {17} (\bibinfo {year} {1993})}\BibitemShut
  {NoStop}%
\bibitem [{\citenamefont {Ito}\ \emph {et~al.}(1999)\citenamefont {Ito},
  \citenamefont {Shen}, \citenamefont {Hirth},\ and\ \citenamefont
  {Wolfe}}]{Ito1999}%
  \BibitemOpen
  \bibfield  {author} {\bibinfo {author} {\bibfnamefont {G.}~\bibnamefont
  {Ito}}, \bibinfo {author} {\bibfnamefont {Y.}~\bibnamefont {Shen}}, \bibinfo
  {author} {\bibfnamefont {G.}~\bibnamefont {Hirth}}, \ and\ \bibinfo {author}
  {\bibfnamefont {C.~J.}\ \bibnamefont {Wolfe}},\ }\href {\doibase
  10.1016/S0012-821X(98)00216-7} {\bibfield  {journal} {\bibinfo  {journal}
  {Earth Planet. Sci. Lett.}\ }\textbf {\bibinfo {volume} {165}},\ \bibinfo
  {pages} {81} (\bibinfo {year} {1999})}\BibitemShut {NoStop}%
\bibitem [{\citenamefont {Maclennan}\ \emph {et~al.}(2001)\citenamefont
  {Maclennan}, \citenamefont {Mc{}Kenzie},\ and\ \citenamefont
  {Gronv{\"o}ld}}]{Maclennan2001}%
  \BibitemOpen
  \bibfield  {author} {\bibinfo {author} {\bibfnamefont {J.}~\bibnamefont
  {Maclennan}}, \bibinfo {author} {\bibfnamefont {D.}~\bibnamefont
  {Mc{}Kenzie}}, \ and\ \bibinfo {author} {\bibfnamefont {K.}~\bibnamefont
  {Gronv{\"o}ld}},\ }\href {\doibase 10.1016/S0012-821X(01)00553-2} {\bibfield
  {journal} {\bibinfo  {journal} {Earth Planet. Sci. Lett.}\ }\textbf {\bibinfo
  {volume} {194}},\ \bibinfo {pages} {67} (\bibinfo {year} {2001})}\BibitemShut
  {NoStop}%
\bibitem [{\citenamefont {Dalton}\ \emph {et~al.}(2014)\citenamefont {Dalton},
  \citenamefont {Langmuir},\ and\ \citenamefont {Gale}}]{Dalton2014}%
  \BibitemOpen
  \bibfield  {author} {\bibinfo {author} {\bibfnamefont {C.~A.}\ \bibnamefont
  {Dalton}}, \bibinfo {author} {\bibfnamefont {C.~H.}\ \bibnamefont
  {Langmuir}}, \ and\ \bibinfo {author} {\bibfnamefont {A.}~\bibnamefont
  {Gale}},\ }\href {\doibase 10.1126/science.1249466} {\bibfield  {journal}
  {\bibinfo  {journal} {Science}\ }\textbf {\bibinfo {volume} {344}},\ \bibinfo
  {pages} {80} (\bibinfo {year} {2014})}\BibitemShut {NoStop}%
\bibitem [{\citenamefont {Jenkins}\ \emph {et~al.}(2018)\citenamefont
  {Jenkins}, \citenamefont {Maclennan}, \citenamefont {Green}, \citenamefont
  {Cottaar}, \citenamefont {Deuss},\ and\ \citenamefont {White}}]{jenkins18}%
  \BibitemOpen
  \bibfield  {author} {\bibinfo {author} {\bibfnamefont {J.}~\bibnamefont
  {Jenkins}}, \bibinfo {author} {\bibfnamefont {J.}~\bibnamefont {Maclennan}},
  \bibinfo {author} {\bibfnamefont {R.~G.}\ \bibnamefont {Green}}, \bibinfo
  {author} {\bibfnamefont {S.}~\bibnamefont {Cottaar}}, \bibinfo {author}
  {\bibfnamefont {A.~F.}\ \bibnamefont {Deuss}}, \ and\ \bibinfo {author}
  {\bibfnamefont {R.~S.}\ \bibnamefont {White}},\ }\href {\doibase
  10.1029/2017JB015121} {\bibfield  {journal} {\bibinfo  {journal} {J. Geophys.
  Res. -- Solid Earth}\ }\textbf {\bibinfo {volume} {123}},\ \bibinfo {pages}
  {5190} (\bibinfo {year} {2018})}\BibitemShut {NoStop}%
\bibitem [{\citenamefont {Sparks}\ and\ \citenamefont
  {Parmentier}(1991)}]{sparks91}%
  \BibitemOpen
  \bibfield  {author} {\bibinfo {author} {\bibfnamefont {D.}~\bibnamefont
  {Sparks}}\ and\ \bibinfo {author} {\bibfnamefont {E.}~\bibnamefont
  {Parmentier}},\ }\href {\doibase 10.1016/0012-821X(91)90178-K} {\bibfield
  {journal} {\bibinfo  {journal} {Earth Planet. Sci. Lett.}\ }\textbf {\bibinfo
  {volume} {105}} (\bibinfo {year} {1991}),\
  10.1016/0012-821X(91)90178-K}\BibitemShut {NoStop}%
\bibitem [{\citenamefont {Kelemen}\ \emph {et~al.}(1995)\citenamefont
  {Kelemen}, \citenamefont {Shimizu},\ and\ \citenamefont
  {Salters}}]{kelemen95}%
  \BibitemOpen
  \bibfield  {author} {\bibinfo {author} {\bibfnamefont {P.~B.}\ \bibnamefont
  {Kelemen}}, \bibinfo {author} {\bibfnamefont {N.}~\bibnamefont {Shimizu}}, \
  and\ \bibinfo {author} {\bibfnamefont {V.~J.~M.}\ \bibnamefont {Salters}},\
  }\href {\doibase 10.1038/375747a0} {\bibfield  {journal} {\bibinfo  {journal}
  {Nature}\ }\textbf {\bibinfo {volume} {375}},\ \bibinfo {pages} {747}
  (\bibinfo {year} {1995})}\BibitemShut {NoStop}%
\bibitem [{\citenamefont {Kelemen}\ \emph {et~al.}(2000)\citenamefont
  {Kelemen}, \citenamefont {Braun},\ and\ \citenamefont {Hirth}}]{kelemen00}%
  \BibitemOpen
  \bibfield  {author} {\bibinfo {author} {\bibfnamefont {P.~B.}\ \bibnamefont
  {Kelemen}}, \bibinfo {author} {\bibfnamefont {M.}~\bibnamefont {Braun}}, \
  and\ \bibinfo {author} {\bibfnamefont {G.}~\bibnamefont {Hirth}},\ }\href
  {\doibase 10.1029/1999GC000012} {\bibfield  {journal} {\bibinfo  {journal}
  {Geochem.\ Geophys.\ Geosys.}\ }\textbf {\bibinfo {volume} {1}} (\bibinfo
  {year} {2000}),\ 10.1029/1999GC000012}\BibitemShut {NoStop}%
\bibitem [{\citenamefont {Liang}\ \emph {et~al.}(2010)\citenamefont {Liang},
  \citenamefont {Schiemenz}, \citenamefont {Hesse}, \citenamefont
  {Parmentier},\ and\ \citenamefont {Hesthaven}}]{liang10}%
  \BibitemOpen
  \bibfield  {author} {\bibinfo {author} {\bibfnamefont {Y.}~\bibnamefont
  {Liang}}, \bibinfo {author} {\bibfnamefont {A.}~\bibnamefont {Schiemenz}},
  \bibinfo {author} {\bibfnamefont {M.~A.}\ \bibnamefont {Hesse}}, \bibinfo
  {author} {\bibfnamefont {E.~M.}\ \bibnamefont {Parmentier}}, \ and\ \bibinfo
  {author} {\bibfnamefont {J.~S.}\ \bibnamefont {Hesthaven}},\ }\href {\doibase
  10.1029/2010GL044162} {\bibfield  {journal} {\bibinfo  {journal} {Geophys.\
  Res.\ Lett.}\ }\textbf {\bibinfo {volume} {37}} (\bibinfo {year} {2010}),\
  10.1029/2010GL044162}\BibitemShut {NoStop}%
\bibitem [{\citenamefont {Pec}\ \emph {et~al.}(2015)\citenamefont {Pec},
  \citenamefont {Holtzman}, \citenamefont {Zimmerman},\ and\ \citenamefont
  {Kohlstedt}}]{pec15}%
  \BibitemOpen
  \bibfield  {author} {\bibinfo {author} {\bibfnamefont {M.}~\bibnamefont
  {Pec}}, \bibinfo {author} {\bibfnamefont {B.~K.}\ \bibnamefont {Holtzman}},
  \bibinfo {author} {\bibfnamefont {M.~E.}\ \bibnamefont {Zimmerman}}, \ and\
  \bibinfo {author} {\bibfnamefont {D.~L.}\ \bibnamefont {Kohlstedt}},\ }\href
  {\doibase 10.1130/G36611.1} {\bibfield  {journal} {\bibinfo  {journal}
  {Geology}\ }\textbf {\bibinfo {volume} {43}},\ \bibinfo {pages} {575}
  (\bibinfo {year} {2015})}\BibitemShut {NoStop}%
\bibitem [{\citenamefont {Pec}\ \emph {et~al.}(2017)\citenamefont {Pec},
  \citenamefont {Holtzman}, \citenamefont {Zimmerman},\ and\ \citenamefont
  {Kohlstedt}}]{pec17}%
  \BibitemOpen
  \bibfield  {author} {\bibinfo {author} {\bibfnamefont {M.}~\bibnamefont
  {Pec}}, \bibinfo {author} {\bibfnamefont {B.~K.}\ \bibnamefont {Holtzman}},
  \bibinfo {author} {\bibfnamefont {M.~E.}\ \bibnamefont {Zimmerman}}, \ and\
  \bibinfo {author} {\bibfnamefont {D.~L.}\ \bibnamefont {Kohlstedt}},\ }\href
  {\doibase 10.1093/petrology/egx043} {\bibfield  {journal} {\bibinfo
  {journal} {J. Petrol.}\ }\textbf {\bibinfo {volume} {58}},\ \bibinfo {pages}
  {979} (\bibinfo {year} {2017})}\BibitemShut {NoStop}%
\bibitem [{\citenamefont {Keller}\ \emph {et~al.}(2017)\citenamefont {Keller},
  \citenamefont {Katz},\ and\ \citenamefont {Hirschmann}}]{keller2017}%
  \BibitemOpen
  \bibfield  {author} {\bibinfo {author} {\bibfnamefont {T.}~\bibnamefont
  {Keller}}, \bibinfo {author} {\bibfnamefont {R.~F.}\ \bibnamefont {Katz}}, \
  and\ \bibinfo {author} {\bibfnamefont {M.~M.}\ \bibnamefont {Hirschmann}},\
  }\href {\doibase 10.1016/j.epsl.2017.02.006} {\bibfield  {journal} {\bibinfo
  {journal} {Earth Planet. Sci. Lett.}\ }\textbf {\bibinfo {volume} {464}},\
  \bibinfo {pages} {55} (\bibinfo {year} {2017})}\BibitemShut {NoStop}%
\bibitem [{\citenamefont {Rees~Jones}\ and\ \citenamefont
  {Katz}(2018)}]{reesjones2018-jfm}%
  \BibitemOpen
  \bibfield  {author} {\bibinfo {author} {\bibfnamefont {D.~W.}\ \bibnamefont
  {Rees~Jones}}\ and\ \bibinfo {author} {\bibfnamefont {R.~F.}\ \bibnamefont
  {Katz}},\ }\href {\doibase 10.1017/jfm.2018.524} {\bibfield  {journal}
  {\bibinfo  {journal} {J. Fluid Mech.}\ }\textbf {\bibinfo {volume} {852}},\
  \bibinfo {pages} {5} (\bibinfo {year} {2018})}\BibitemShut {NoStop}%
\bibitem [{\citenamefont {Kelemen}\ \emph {et~al.}(1997)\citenamefont
  {Kelemen}, \citenamefont {Hirth}, \citenamefont {Shimizu}, \citenamefont
  {Spiegelman},\ and\ \citenamefont {Dick}}]{kelemen97}%
  \BibitemOpen
  \bibfield  {author} {\bibinfo {author} {\bibfnamefont {P.~B.}\ \bibnamefont
  {Kelemen}}, \bibinfo {author} {\bibfnamefont {G.}~\bibnamefont {Hirth}},
  \bibinfo {author} {\bibfnamefont {N.}~\bibnamefont {Shimizu}}, \bibinfo
  {author} {\bibfnamefont {M.}~\bibnamefont {Spiegelman}}, \ and\ \bibinfo
  {author} {\bibfnamefont {H.~J.~B.}\ \bibnamefont {Dick}},\ }\href {\doibase
  10.1098/rsta.1997.0010} {\bibfield  {journal} {\bibinfo  {journal} {Phil.\
  Trans.\ R.\ Soc.\ London A}\ }\textbf {\bibinfo {volume} {355}},\ \bibinfo
  {pages} {283} (\bibinfo {year} {1997})}\BibitemShut {NoStop}%
\bibitem [{\citenamefont {Gee}\ \emph {et~al.}(1998)\citenamefont {Gee},
  \citenamefont {Thirlwall}, \citenamefont {Taylor}, \citenamefont {Lowry},\
  and\ \citenamefont {Murton}}]{Gee1998-JPet}%
  \BibitemOpen
  \bibfield  {author} {\bibinfo {author} {\bibfnamefont {M.~A.~M.}\
  \bibnamefont {Gee}}, \bibinfo {author} {\bibfnamefont {M.~F.}\ \bibnamefont
  {Thirlwall}}, \bibinfo {author} {\bibfnamefont {R.~N.}\ \bibnamefont
  {Taylor}}, \bibinfo {author} {\bibfnamefont {D.}~\bibnamefont {Lowry}}, \
  and\ \bibinfo {author} {\bibfnamefont {B.~J.}\ \bibnamefont {Murton}},\
  }\href {\doibase 10.1093/petroj/39.5.819} {\bibfield  {journal} {\bibinfo
  {journal} {J. Petrol.}\ }\textbf {\bibinfo {volume} {39}},\ \bibinfo {pages}
  {819} (\bibinfo {year} {1998})}\BibitemShut {NoStop}%
\bibitem [{\citenamefont {Rees~Jones}\ and\ \citenamefont
  {Rudge}(2019)}]{ReesJones19Code}%
  \BibitemOpen
  \bibfield  {author} {\bibinfo {author} {\bibfnamefont {D.~W.}\ \bibnamefont
  {Rees~Jones}}\ and\ \bibinfo {author} {\bibfnamefont {J.~F.}\ \bibnamefont
  {Rudge}},\ }\href {\doibase 10.5281/zenodo.3524067} {\bibfield  {journal}
  {\bibinfo  {journal} {Github}\ } (\bibinfo {year} {2019}),\
  10.5281/zenodo.3524067}\BibitemShut {NoStop}%
\end{thebibliography}%

\vspace*{1.0cm}

\noindent {\centering \textbf{{\large{SUPPLEMENTARY MATERIAL}} } }
\setcounter{section}{0}
\setcounter{figure}{0}
\setcounter{equation}{0}

\renewcommand\thefigure{S\arabic{figure}}  
\renewcommand\thesection{S\arabic{section}}  
\renewcommand\theequation{S\arabic{equation}} 

\vspace*{0.2 cm}

\textbf{Solution method for time-dependent equations}

\vspace*{0.2 cm}

In the main text, we derived a simplified set of equations (\ref{eq:transient-scaled}, \ref{eq:Q-phi-transient-n2}) for the time-dependent porosity and melt flux. 
These are (considering the case $n=2$):
\begin{equation} \label{S-eq:transient-scaled}
\frac{\partial \hat{\phi}}{\partial \hat{t}} + \frac{\partial \hat{Q} }{\partial \hat{z}} = f(\hat{t}) ,
\end{equation}
\begin{equation} \label{S-eq:Q-phi-transient-n2}
\hat{Q} = 2 \hat{z}^{1/2} \hat{\phi} + \mathcal{A} \hat{\phi}^2,
\end{equation}
where $f(\hat{t})=1$ for $0\leq\hat{t}\leq \lambda$ and zero outside this interval.

\subsection{Method of characteristics}

\begin{figure}
\noindent\includegraphics[width=0.85\linewidth]{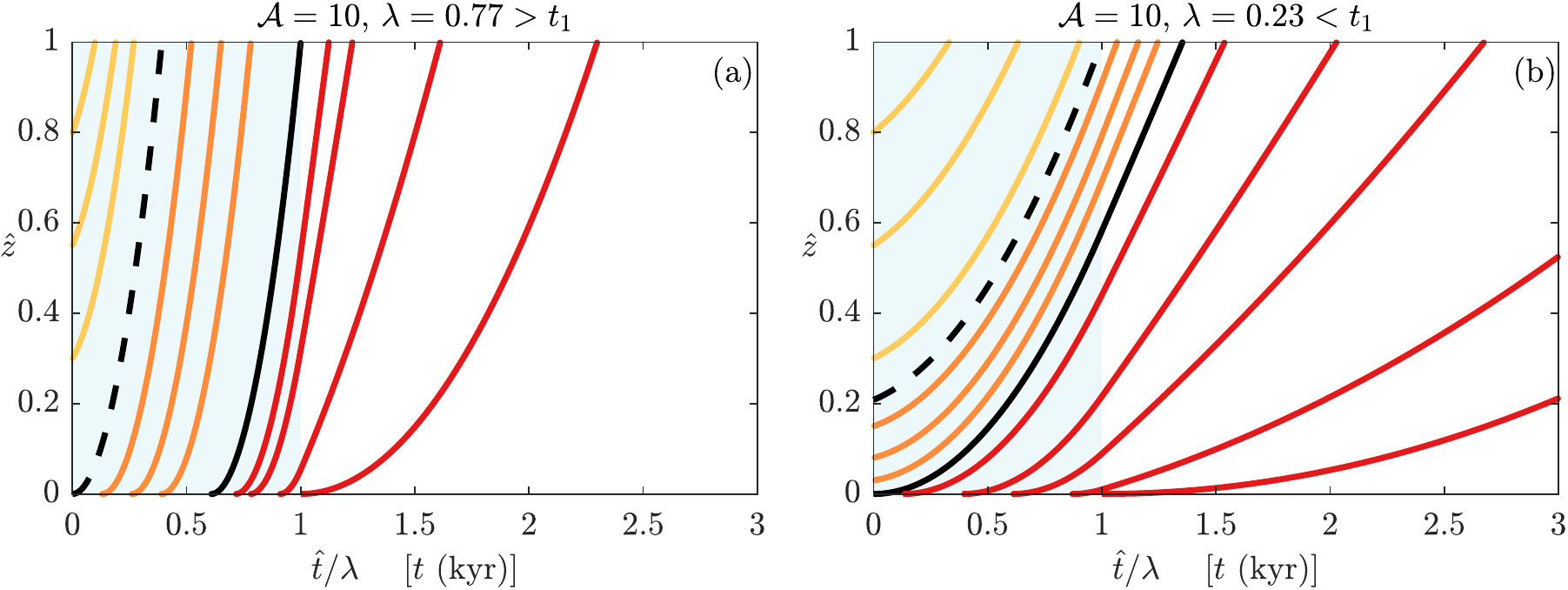}
\caption{Examples showing sets of characteristics for the two cases: (a) $ \lambda>\hat{t}_1$  and (b) $ \lambda<\hat{t}_1$. The parameter values chosen correspond to the examples from figure~\ref{fig:timeseries} of intermediate nonlinearity ($\mathcal{A}=10$) and medium and slow melt velocities. In particular, $\lambda= 0.77$ is equivalent to $\overline{w}_\mathrm{max} = 100$~m/yr and $\lambda=0.23$ is equivalent to $\overline{w}_\mathrm{max} = 30$~m/yr.
Yellow curves correspond to the first phase, when emissions are rising.
Orange curves correspond to the second phase, when emissions are flatlining or falling slightly.  
Red curves correspond to the final phase, when emissions are falling to zero. Black curves highlight transitions between these phases.  }
\label{fig:characteristics}
\end{figure}

This is a hyperbolic system of equations, so a good analytical method to solve it is the method of characteristics. 
Figure~\ref{fig:characteristics} shows two sets of characteristics.
The idea is to introduce a characteristic curve $[\hat{z}(\hat{s}),\hat{t}(\hat{s})]$, parameterized by a new variable $\hat{s}$.
We then consider the evolution of $\hat{Q}(\hat{s})=\hat{Q}[\hat{s},\hat{t}(\hat{s})]$, i.e. to the evolution of $\hat{Q}$ along the characteristic curve.
Then
\begin{equation} \label{S-eq:dQds1}
\frac{d}{d\hat{s}} \hat{Q}(\hat{s}) = \frac{\partial \hat{Q}}{\partial \hat{t}}\frac{d\hat{t}}{d\hat{s}} + \frac{\partial \hat{Q} }{\partial \hat{z}} \frac{d\hat{z}}{d\hat{s}}.
\end{equation}
Now equation~(\ref{S-eq:Q-phi-transient-n2}) is a quadratic equation in $\hat{\phi}$ so can be rearranged to give $\hat{\phi}(\hat{Q})$:
\begin{equation} \label{S-eq:phi(Q)}
\hat{\phi} ={\mathcal{A}}^{-1} \left[{-\hat{z}^{1/2}+(z+\mathcal{A} \hat{Q} )^{1/2} }\right]. 
\end{equation}
Equation~(\ref{S-eq:Q-phi-transient-n2}) can also be differentiated, and combining with equation~(\ref{S-eq:phi(Q)}), we find
\begin{equation} \label{S-eq:dQdt1}
 \frac{\partial \hat{Q}}{\partial \hat{t}} = \frac{\partial \hat{\phi}}{\partial \hat{t}} \left(2\hat{z}^{1/2} + 2\mathcal{A} \hat{\phi} \right) =  \frac{\partial \hat{\phi}}{\partial \hat{t}} \, 2  \left(\hat{z} + \mathcal{A} \hat{Q} \right)^{1/2} . 
\end{equation} 
So if we let
\begin{equation}
\frac{d\hat{t}}{d\hat{s}} = 1
\end{equation} 
and
\begin{equation}  \label{S-eq:z_char}
\frac{d\hat{z}}{d\hat{s}} = 2  \left(\hat{z} + \mathcal{A} \hat{Q} \right)^{1/2} ,
\end{equation} 
then by combining equations~(\ref{S-eq:dQds1}, \ref{S-eq:dQdt1}), we find
\begin{equation} \label{S-eq:dQds2}
\frac{d}{d\hat{s}} \hat{Q}(\hat{s}) =f \frac{d\hat{z}}{d\hat{s}}.
\end{equation}
Inside the forcing region $f=1$, so 
\begin{equation} \label{S-eq:Q_char}
\hat{Q}-\hat{Q}_0 = \hat{z}-\hat{z}_0,
\end{equation}
where $\hat{Q}_0=0$ is the initial/boundary condition (at the start of a characteristic).
Outside the forcing region, $f=0$, $d\hat{Q}=0$ (i.e. $\hat{Q}$ is constant along a characteristic outside the forcing region). 
Then we back substitute $\hat{Q}$ into equation~(\ref{S-eq:z_char}) and then integrate (by hand) to obtain $\hat{z}(\hat{s})$. 
Note that $\hat{t} = \hat{s}$, so we have the characteristic and the flux at every point along it. 
Finally, we evaluate the flux at the end of characteristic \mbox{($\hat{z}=1$)}, which is the top of the melting column.

To illustrate the method, consider a characteristic starting at $\hat{z}_0$ at $\hat{t}=0$. 
The following example corresponds to the yellow curves in figure~\ref{fig:characteristics}.
Provided the characteristic remains in the forcing window,  $\hat{Q}  = \hat{z}-\hat{z}_0$ and so
\begin{equation}  \label{S-eq:z_char-eg1}
\frac{d\hat{z}}{d\hat{s}} = 2  \left[(1 + \mathcal{A})\hat{z} - \mathcal{A} \hat{z}_0  \right]^{1/2} ,
\end{equation} 
This is a separable equation so can be integrated to find
\begin{equation} \label{S-eq:char_regime1}
\hat{s}=\frac{1}{1+\mathcal{A}} \left\{ \left[ (1 + \mathcal{A})\hat{z} - \mathcal{A} \hat{z}_0\right]^{1/2}  - \hat{z}_0^{1/2} \right\},
\end{equation}
applying the initial condition.
Then the flux at the top $\hat{Q}(\hat{z}=1,\hat{t})$ satisfies
\begin{equation}
\hat{t}=\frac{1}{1+\mathcal{A}} \left\{ \left( 1 + \mathcal{A}\hat{Q}  \right)^{1/2}  - (1-\hat{Q})^{1/2} \right\}.
\end{equation}
This is a quadratic equation for $\hat{Q}$ and can be rearranged to give 
\begin{equation} \label{S-eq:Q_regime1}
\hat{Q}= 2 \hat{t} \left(1-\hat{t}^2\mathcal{A} \right)^{1/2} - \hat{t}^2(1-\mathcal{A}).
\end{equation}
This equation holds for $0\leq \hat{t} \leq \lambda$ for the case $\lambda\leq \hat{t}_1\equiv (1+\mathcal{A})^{-1/2}$ or for  $0\leq \hat{t} \leq \hat{t}_1$ for the case $\lambda\geq \hat{t}_1$.
Physically, the rise in emissions continues either until the end of the deglaciation period, or until a parcel of extra melt travels across the entire melting region (which takes a time $\hat{t}_1$).

We then apply the same method for the whole period, taking into account when characteristics leave the forcing window, as illustrated in figure~\ref{fig:characteristics}. 
We describe the results more briefly, going through the two cases (Appendix A of the main text) separately. 
We divide up periods according to the time a characteristic reaches the top of the melting column.

\subsection{Case 1: $ \lambda>\hat{t}_1$}
For this case melt transport is relatively fast. 

\subsubsection{Period 1: $0\leq \hat{t} \leq \hat{t}_1$}
For the first period, the characteristics are described by equation~(\ref{S-eq:char_regime1}) and the flux at the top by equation~(\ref{S-eq:Q_regime1}).
Example characteristics are plotted in yellow in figure~\ref{fig:characteristics}(a) and the dashed black curve that reaches the top at $\hat{t}=\hat{t}_1$ marks the end of this period.
Thus $\hat{t}_1$ corresponds to the time it takes for a parcel of extra melt to travel from the bottom to the top.
The significance of the criterion $ \lambda>\hat{t}_1$ is that melt transport is sufficiently fast that this characteristic reaches the top before deglaciation finishes.

\subsubsection{Period 2: $\hat{t}_1\leq \hat{t} \leq \lambda$}
For the second period, characteristics that start at $\hat{t}=\hat{t}_0$ are described by 
\begin{equation} \label{S-eq:char_regime2A}
\hat{s}= \hat{t}_0 + \left( \frac{\hat{z}}{1 + \mathcal{A}} \right)^{1/2},
\end{equation}
and the flux at the top
\begin{equation} \label{S-eq:Q_regime2A}
\hat{Q}= 1.
\end{equation}
Example characteristics are plotted in orange in figure~\ref{fig:characteristics}(a) and the solid black curve that reaches the top at $\hat{t}=\lambda$ marks the end of this period.
Note that all the characteristics are the same in this period (only offset in time).
They all accumulate the same amount of melt, giving rise to equation~(\ref{S-eq:Q_regime2A}).

\subsubsection{Period 3: $\lambda \leq \hat{t} \leq \lambda + 1$}
For the third period, characteristics are initially described by the same equation as before,~(\ref{S-eq:char_regime2A}).
However, this ceases to apply after $s=\lambda$, or equivalently above $\hat{z}_\lambda$, where
\begin{equation}
\hat{z}_\lambda = (1+\mathcal{A})(\lambda-\hat{t}_0)^2.
\end{equation}
Thereafter, $\hat{Q}=\hat{Q}_\lambda\equiv \hat{z}_\lambda$ and the characteristics are given by
\begin{equation} \label{S-eq:char_regime3A}
\hat{s}= \lambda + \left( \hat{z} + \mathcal{A} \hat{Q}_\lambda \right)^{1/2} -(1+\mathcal{A})(\lambda-\hat{t}_0).
\end{equation}
The flux at the top is given by
\begin{equation} \label{S-eq:Q_regime3A}
\hat{Q}= 1+ (\hat{t}-\lambda)^2 (1+2\mathcal{A}) - 2(1+\mathcal{A})^{1/2}(\hat{t}-\lambda)\left[1+\mathcal{A}(\hat{t}-\lambda)^2\right]^{1/2}.
\end{equation}
Example characteristics are plotted in red in figure~\ref{fig:characteristics}(a) and the final red characteristics that reaches the top at $\hat{t}=\lambda+1$ marks the end of this period.

\subsubsection{Cumulative emissions}
The cumulative emissions are obtained by integrating with respect to time.
We find
\begin{equation} 
    \lambda \hat{F} = \left\{\begin{array}{ll}
         \hat{C}_1 & (0\leq \hat{t} \leq \hat{t}_1),\\
         \hat{C}_2 & (\hat{t}_1\leq \hat{t} \leq \lambda),\\
         \hat{C}_3 & (\lambda \leq \hat{t} \leq 1+\lambda),
        \end{array}\right.
\end{equation} 
where
\begin{equation} \label{eq:S-eq:cumulative_1A}
\hat{C}_1 = \frac{2}{3\mathcal{A}} \left[ 1-(1-\hat{t}^2\mathcal{A})^{3/2} \right]-\frac{\hat{t}^3}{3}(1-\mathcal{A}), 
\end{equation}
\begin{equation}
\hat{C}_2 =\hat{t}+  \frac{2}{3\mathcal{A}} \left[ 1-(1+\mathcal{A})^{1/2}\right],
\end{equation}
and
\begin{equation}
\hat{C}_3 =\hat{t}+ \frac{(\hat{t}-\lambda)^3}{3}(1+2 \mathcal{A}) + \frac{2}{3\mathcal{A}} \left\{ 1-(1+\mathcal{A})^{1/2} \left[ 1+\mathcal{A}(\hat{t}-\lambda)^2 \right]^{3/2} \right\}.
\end{equation}

\subsection{Case 2: $ \lambda<\hat{t}_1$}
This is the case of relatively slow melt transport. 
\subsubsection{Period 1: $0\leq \hat{t} \leq \lambda$}
As before, for the first period, the characteristics are described by equation~(\ref{S-eq:char_regime1}) and the flux at the top by equation~(\ref{S-eq:Q_regime1}).
Example characteristics are plotted in yellow in figure~\ref{fig:characteristics}(b) and the dashed black curve that reaches the top at $\hat{t}=\lambda$ marks the end of this period.
Characteristics that arrive later only arrive after deglaciation has finished.

\subsubsection{Period 2: $\lambda\leq \hat{t} \leq \hat{t}_2$}
For the second period, characteristics are again described by equation~(\ref{S-eq:char_regime1}).
However, this 
ceases to apply after $s=\lambda$, or equivalently above $\hat{z}_\lambda$, where
\begin{equation}
\hat{z}_\lambda = \frac{1}{1+\mathcal{A}}  \left(\mathcal{A} \hat{z}_0+\left[(1+\mathcal{A})\lambda+\hat{z}_0^{1/2}\right]^2 \right) .
\end{equation}
Thereafter, $\hat{Q}=\hat{Q}_\lambda\equiv \hat{z}_\lambda-\hat{z}_0$ and the characteristics are given by 
\begin{equation} \label{S-eq:char_regime2B}
\hat{s}=\lambda+\left(\hat{z}+ \mathcal{A} \hat{Q}_\lambda\right)^{1/2}-\left(\hat{z}_\lambda+\mathcal{A}  \hat{Q}_\lambda\right)^{1/2} 
\end{equation}
and the flux at the top
\begin{equation} \label{S-eq:Q_regime2B}
\hat{Q}=  \lambda \left\{-2\hat{t} +\lambda(1+\mathcal{A}) +2 \left[1+\mathcal{A} \lambda (\lambda-2\hat{t} )\right]^{1/2} \right\}.
\end{equation}
Example characteristics are plotted in orange in figure~\ref{fig:characteristics}(b) and the solid black curve that reaches the top at $\hat{t}=\hat{t}_2$ marks the end of this period.
Thus $\hat{t}_2$ is the travel time from bottom to top for a parcel of extra melt that starts at the bottom at $t=0$ and reaches the top after deglaciation has finished.
The nonlinear nature of the system means that the travel time depends on the starting time for a parcel that reaches the top after deglaciation has finished.

\subsubsection{Period 3: $\hat{t}_2 \leq \hat{t} \leq \lambda + 1$}
For the third period, is exactly the same as for case 1. 
Example characteristics are plotted in red in figure~\ref{fig:characteristics}(b) and the final red characteristics that reaches the top at $\hat{t}=\lambda+1$ marks the end of this period.

\subsubsection{Cumulative emissions}
Again, the cumulative emissions are obtained by integrating with respect to time.
We find
\begin{equation} 
    \lambda \hat{F} = \left\{\begin{array}{ll}
         \hat{C}_1 & (0\leq \hat{t} \leq \lambda),\\
         \hat{C}_2 & (\lambda \leq \hat{t} \leq \hat{t}_2),\\
         \hat{C}_3 & (\hat{t}_2 \leq \hat{t} \leq 1+\lambda),
        \end{array}\right.
\end{equation} 
where $\hat{C}_1$ is the same as in case 1, given by equation~(\ref{eq:S-eq:cumulative_1A}), 
\begin{equation}
\hat{C}_2 =\lambda \left[ (\hat{t}-\lambda)(\lambda\mathcal{A}-\hat{t})- \frac{\lambda^2}{3}(1-\mathcal{A}) \right] +\frac{2}{3 \mathcal{A}} \left[1-(1+\lambda\mathcal{A}(\lambda-2\hat{t}))^{3/2} \right]
\end{equation}
and
\begin{eqnarray}
\hat{C}_3 &=& \lambda \left[(\hat{t}_2-\lambda)(\lambda\mathcal{A}-\hat{t}_2)-\frac{\lambda^2}{3}(1-\mathcal{A}) \right]+\frac{2}{3\mathcal{A}} \left[1-(1+\lambda\mathcal{A}(\lambda- 2 \hat{t}_2))^{3/2})\right]  \nonumber \\
&+ &                 \hat{t}-\hat{t}_2+\frac{1+2\mathcal{A}}{3} \left[ (\hat{t}-\lambda)^3-(\hat{t}_2-\lambda)^3 \right] \nonumber \\
&- &    \frac{2}{3 \mathcal{A}} (1+\mathcal{A})^{1/2}  \left\{ \left[1+\mathcal{A} (\hat{t}-\lambda)^2\right]^{3/2}- \left[1+\mathcal{A}(\hat{t}_2-\lambda)^2 \right]^{3/2} \right\}.
\end{eqnarray}
The first line in the expression for $\hat{C}_3$ is $\hat{C}_2(\hat{t}_2)$.     
    
\end{document}